# Formation of Trans-Neptunian Satellite Systems at the Stage of Condensations

S. I. Ipatov[a,b]

[a]*Vernadsky Institute of Geochemistry and Analytical Chemistry,
Russian Academy of Sciences, Moscow, Russia*
[b]*Space Research Institute, Russian Academy of Sciences, Moscow, Russia*
e-mail: siipatov@hotmail.com

**Abstract**—The formation of trans-Neptunian satellite systems at the stage of rarefied preplanetesimals (i.e., condensations of dust and/or objects less than 1 m in diameter) is discussed. It is assumed that trans-Neptunian objects (including those with satellites) could form as a result of compression of parental rarefied pre-planetesimals. The formulas for calculating the angular momentum of two colliding condensations with respect to their center of mass, which were applied earlier in (Ipatov, 2010) in the comparison of such momenta with the angular momenta of observed satellite systems, are used to estimate the angular momenta of condensations needed to form satellite systems. It is demonstrated that the angular velocities of condensations used in (Nesvorny et al., 2010) as the initial data in the computer simulation of compression of rarefied preplanetesimals and subsequent formation of trans-Neptunian satellite systems may be obtained in collisions of preplanetesimals with their radii comparable to the corresponding Hill radii. For example, these angular velocities are in the range of possible values of angular velocities of a parental rarefied preplanetesimal formed as a result of a merger of two colliding rarefied preplanetesimals that moved in circular heliocentric orbits before a collision. Some rarefied preplanetesimals formed as a result of collision of preplanetesimals in the region of formation of solid small bodies acquire such angular momenta that are sufficient to form satellite systems of small bodies. It is likely that the ratio of the number of rarefied preplanetesimals with such angular momenta to the total number of rarefied preplanetesimals producing classical trans-Neptunian objects with diameters larger than 100 km was 0.45 (the initial fraction of satellite systems among all classical trans-Neptunian objects).

## INTRODUCTION

### Formation of Preplanetesimals and Planetesimals

Models of formation of solid planetesimals from rarefied preplanetesimals have been discussed since the 1950s (e.g., Safronov, 1972; Goldreich and Ward, 1973; Vityazev et al., 1990). Such condensed regions of a protoplanetary disk are bound gravitationally with each other and contain solid particles and gas. These regions were sometimes called condensations. Models of planetesimal accretion from small solid objects (e.g., Weidenschilling, 2003) became popular in the 1990s, since it was assumed that self-generating turbulence in the central plane interferes with the gravitational instability of rarefied preplanetesimals. New arguments for the formation of rarefied preplanetesimals (condensations) were found in the 2000s (see, for example, Makalkin and Ziglina, 2004; Johansen et al., 2007; 2009a; 2009b; 2011; 2012; 2015a; 2015b; Cuzzi et al., 2008; 2010; Lyra et al., 2008; 2009; Chambers, 2010; Chiang and Youdin, 2010; Rein et al., 2010; Youdin, 2011; Youdin and Kenyon, 2013; Marov et al., 2013; Wahlberg Jansson and Johansen, 2014; Carrera et al., 2015; Kretke and Levison, 2015; Wahlberg Jans- son et al., 2017; Ziglina and Makalkin, 2016). In contrast to purely dust condensations considered earlier, these condensations could contain solid objects (decimeter- or meter-sized boulders and smaller pebbles). In what follows, any condensations containing dust and/or objects less than 1 m in diameter are called rarefied preplanetesimals. The term "preplanetesimals" is often used for brevity.

It was concluded in (Makalkin and Ziglina, 2004) that the dust subdisk could reach critical density in the trans-Neptunian region (at a distance of more than 30 AU from the Sun), since (in contrast to the region of planet formation) shear turbulence did not penetrate through to the equatorial plane here. Therefore, the inner (equatorial) subdisk layer, which, unlike two surface layers, was not subject to shear turbulence, could be gravitationally unstable. It was suggested that this layer broke down into condensations with a size of $\sim 10^{12}$ cm at 40 AU from the Sun. Small fragments ($\sim 10^9$ cm) of these condensations could compress rapidly and form bodies with a diameter of $\sim 10$ km. In $\sim 10^6$ years, larger ($\sim 100$ km) planetesimals formed from these bodies with a certain probability. Marov et al. (2013) have analyzed the mechanisms of formation and growth of dust condensations by dust absorption and concluded that planetesimals with a mass of $\sim 10^{21}$–$10^{22}$ g and a radius of 50–100 km could form in the terrestrial region in $\sim 10^3$–$10^4$ years. Ziglina and Makalkin (2016) have shown that gravitational insta- bility in the layer at one AU with the current turbu- lence models requires the disk to be enriched with sol- ids greater by a factor of 5–10 than the initial content and particle aggregates to become larger than 0.3 m in size. At a distance of 10 AU, such enrichment and growth are not needed.

Carrera et al. (2015) have found that solid particles with their size ranging from 1 mm to 1 m could form dense clouds owing to a rapidly enhancing radial drift convergence (streaming instability). Specifically, they have noted that streaming instability emerges easily in the case of millimeter particles at distance $a > 10$ AU from the Sun; as particles grow larger, streaming instability becomes more efficient. It was concluded that the particle size of 1 mm is the lower limit for stream- ing instability. Carrera et al. (2015) have assumed that the planetesimals formed via condensation collapse were 100–1000 km in size (both for asteroids and for trans-Neptunian objects). radii ranging from 10 to several thou- sand kilometers in the process of compression of Wahlberg Jansson et al. (2017) have studied the formation of solid planetesi- mals with their condensations comparable in size to the Hill sphere at ~40 AU from the Sun. The typical size of particles in the calculations of streaming instability in (Johansen et al., 2007; 2009b; 2015b) varied from 10 cm to 1 m.

Efficient formation of gravitationally bound condensations with their masses corresponding to the radii of planetesimals (formed via compression of condensations) ranging from 100 to 400 km in the asteroid belt or from 150 to 730 km in the Edgeworth–Kuiper belt was found in (Johansen et al., 2012). The authors noted that their calculations were performed for dimensionless parameters, and the masses of formed planetesimals are proportional to $a^{3/4}$, where $a$ a is the distance from these planetesimals to the Sun. Therefore, the characteristic masses of planetesimals at 30 AU are 5–6 times higher than those in the asteroid belt region. These results agree with the fact that the typical and maximum masses of trans-Neptunian objects are higher than those of asteroids.

Several binary planetesimals were obtained in calculations with the highest resolution in (Johansen et al., 2015b). Arguments in favor of the theory that asteroids were born large (with diameter $d > 100$ km) were examined in (Morbidelli et al., 2009). Davidson et al. (2016) have assumed in their model that trans- Neptunian objects formed via streaming instability were no larger than 400 km, while bigger bodies grew slowly by colliding with neighboring objects.

The results of observations of particle ejection from comets 67P and 103P suggest that comets form via compres- sion of rarefied condensations containing milli-, centi-, and decimeter particles (Gundlach et al., 2015; Kretke and Levison, 2015; Fulle et al., 2016; Poulet et al., 2016).

Streaming instability was not the only mechanism considered to be involved in the formation of rarefied condensations. For example, Johansen et al. (2011) have studied the formation of planetesimals owing to turbulence induced by magnetorotational instability. In their calculations, meter-sized objects concentrated in large regions of excess pressure. The masses of formed planetesimals were as large as several (up to 35) masses of Ceres. Lyra et al. (2008) have examined the instability induced by Rossby waves and considered particles with a diameter of 1–100 cm. Cuzzi et al. (2008; 2010) have studied the formation of dense condensations of millimeter-sized particles. It was concluded that a certain fraction of these dense condensations could undergo compression to the density of solid bodies and form "sandpile" planetesimals (including trans-Neptunian objects) with a typical diameter of 10–100 km. It was found that small bodies dominate the size distribution of planetesimals, but the greater part of mass is contained within several large bodies.

Lambrechts and Johansen (2012) have examined numerically the growth of planetesimals and cores of giant planets by accretion of centimeter-sized particles. The accretion of chondrules onto planetesimal embryos forming via streaming instability was ana- lyzed in (Johansen et al., 2015b). It was concluded that asteroids and certain embryos with radii more than 200 km have acquired at least 2/3 of their mass by chondrule accretion. However, the accretion of milli-meter-sized particles was immaterial at more than 25 AU from the Sun, and planetesimals have retained their initial sizes. At one AU from the Sun, the contribution of chondrules to the growth of planetesimals was small relative to the contribution of preplanetesimals in the first 200 000 years, but became more significant later.

*Times of Formation and Compression of Preplanetesimals*

The number of collisions of preplanetesimals depends on the number of preplanetesimals in the considered region, on their initial sizes, and on the time dependences of radii of contracting preplanetesimals. According to Lyra et al. (2009), Mars-sized preplanetesimals were formed from boulder-sized objects in 40 revolutions around the Sun. The calculations performed by Johansen et al. (2011) have demonstrated that bound clusters form in ten revolutions. In the calculations of Johansen et al. (2012), the maximum ratio of the mass of particles in unit volume to the gas density in the central plane was approximately 3000 and was attained in 25 revolutions around the Sun. The particle concentration did increase by 3–4 orders of magnitude relative to the average value in 15–30 revolutions around the Sun in (Gibbons et al., 2012). Lyra et al. (2008) have reported the formation of more than 300 gravitationally bound embryos (20 of them were more massive than Mars) in 200 orbital periods. The time of compression of condensations needed to form planetesimals larger than 2 km in diameter was no longer than 300 years in the model of Wahlberg Jansson and Johansen (2014).

Chambers (2010) believed that the time of formation of planetesimals from rarefied preplanetesimals decreased (even in years) with distance from the Sun. Cuzzi et al. (2010) have found that certain dense condensations (preplanetesimals) escape disintegration and evolve into dense objects in 100–1000 revolutions around the Sun. Cuzzi and Hogan (2012) have concluded that the rate of planetesimal formation was probably underestimated in (Cuzzi et al., 2010) and overestimated in (Chambers, 2010). Other authors have considered both longer and shorter preplanetesi- mal compression times.

In the calculations performed by Nesvorny et al. (2010), the satellite systems formed from preplanetesimals in 100 years (i.e., in 0.6 of an orbital period at 30 AU) generally contained two or more large objects and hundreds of smaller bodies. Trans-Neptunian objects with a diameter up to 1000 km formed in several million years in the model of Makalkin and Ziglina

(2004); the settlement of dust in the equatorial plane occupied the greater part of this time interval, while it took condensations ~1 million years to compress. Objects with a size of 1000 km formed in the process of compression of clusters of bodies with a diameter of 10 km. Myasnikov and Titarenko (1989a; 1989b) believed that the lifetimes of gas–dust condensations could exceed several million years and depended on the optical properties of the material and on the concentration of short-lived radioactive isotopes.

Wahlberg Jansson and Johansen (2014) have concluded that planetesimals with a radius exceeding 100 km formed via collapse in a free fall time of ~25 years at a distance of 40 AU from the Sun; at smaller masses of condensations, the process of collapse was much more extended. The obtained estimates were so low since the angular momentum of a collapsing condensation was neglected in these calculations. Safronov (1972) and Vityazev et al. (1990) have examined the dependence of the compression time of a rarefied condensation on its angular momentum. Vityazev et al. (1990) have concluded that the condensation compression time was approximately $10^5$–$10^6$ years in the feeding zone of terrestrial planets. According to the estimates obtained by Safronov (1972), the condensation compression time was on the order of $10^4$ years at 1 AU and $10^6$ years at the Sun–Jupiter distance.

The estimated condensation compression times varied widely (from 25 years to millions of years) in the papers cited above. The minimum values of this time corresponded to the free fall model. Longer compression times may be associated with the inclusion of the angular momentum of preplanetesimals and mutual collisions of objects forming these preplanetesimals and with the presence of gas and short-lived radioactive isotopes. The small estimated condensation com- pression times were obtained in relatively simple models that neglect some of the factors influencing the compression of preplanetesimals. The number of objects forming preplanetesimals in the discussed computer models was much lower than the actual number of such objects.

*Formation of Satellite Systems of Small Bodies*

Several hypotheses of the formation of binary small bodies, where collisions and close approaches of solid bodies are considered, are known. For example, Goldreich et al. (2002) have examined the capture of the second component within the Hill sphere via dynamical friction from surrounding small bodies or due to the gravitational influence of the third (large) body. Weidenschilling (2002) has studied a collision of two planetesimals within the sphere of inf luence of the third (more massive) body. Funato et al. (2004) have considered a model in which two bodies first collide and form a binary system with components differing widely in mass. The low-mass second component is then ejected and substituted by the third body in an eccentric orbit. The results obtained in (Astakhov et al., 2005) were based on modeling the four-body prob- lem and included solar tidal effects. Gorkavyi (2008) has proposed a model of multiple collisions. Ćuk (2007), Pravec et al. (2007), and Walsh et al. (2008) have concluded that rotational break-up of a "rubble pile" may be the primary mechanism of formation of binary objects with a low-mass (with a diameter smaller than 10 km) primary object (e.g., near-Earth objects). The angular momentum needed for such break-up is acquired owing to the YORP (Yarkovsky–O'Keefe–Radzievskii–Paddack) effect. Additional references to the hypotheses of formation of binary small bodies may be found in (Richardson and Walsh, 2006; Petit et al., 2008; Noll et al., 2008a). In the case of small bodies with multiple satellites, the largest satellite of the primary component is referred to in the present study when binary small bodies are mentioned.

Unlike the authors of models of formation of binary small bodies (satellite systems) at the stage of solid bodies, Ipatov (2009; 2010; 2014; 2015c) and Nesvorny et al. (2010) have assumed that most large binary trans-Neptunian objects were formed by compression of rarefied preplanetesimals. In our view, the formation of certain other binary objects (especially those with diameter $d < 100$ km of the primary component; e.g., most minor binary objects in the asteroid belt and in the near-Earth population) may be accounted for by certain models that consider solid bodies, which were formed either via preplanetesimal compression or in collisions of larger bodies.

Even before the introduction of new arguments in favor of rarefied preplanetesimals, Ipatov (2001; 2004) has proposed that the majority of planetesimals, trans-Neptunian objects (TNOs), and asteroids with diameter $d > 100$ km could have formed not by accumulation of small solid planetesimals, but as a result of compression of large rarefied preplanetesimals. Direct formation of TNOs of their current sizes from rarefied condensations was contemplated by Eneev (1980). Certain minor objects (TNOs, planetesimals, and asteroids) with $d < 100$ km may be fragments of large objects, while other objects of this kind could form via compression of rarefied preplanetesimals.

Ipatov (2010) has proposed the following mechanisms of formation of a binary object at the stage of compression of a rarefied preplanetesimal that had formed in a collision of two preplanetesimals. (1) In certain cases, a collision of two preplanetesimals results in the formation of a rotating system with two compression centers. The end result of gravitational contraction is a binary system with components located far from each other (e.g., 2001 $QW_{322}$). (2) In other cases, several satellites could form from a disk around the center of the preplanetesimal formed in a collision. If the formed preplanetesimal had acquired an angular momentum exceeding the maximum possible value for a solid body of the same mass, a fraction of its material could escape the preplanetesimal surface in the process of compression and form a cloud that turned into a disk. These two scenarios could coexist.

It was demonstrated in (Ipatov, 2010) that the angular momentum of two collided rarefied preplanetesimals (relative to their center of mass), which moved in circular heliocentric orbits prior to the collision, may correspond to the angular momenta of known trans-Neptunian objects and asteroids with satellites (if the mass of a TNO or an asteroid is the same as the total mass of collided preplanetesimals). The angular momentum of observed binary trans-Neptunian objects varies within a wide range likely since preplanetesimals collided at different stages of their compression (the stronger they are compressed, the lower is the angular momentum acquired in a collision). In addition, the differences in angular momentum could be attributed to variations (with the collision parameters) in the pattern of loss of the angular momentum and mass in the process of compression of the parental condensation or to variations in the difference of semimajor axes of heliocentric orbits of colliding preplanetesimals and the masses of these planetesimals. The effective sizes of collided rarefied preplanetesimals could be comparable to the current distance between the components of a binary system. The angular momenta of certain observed trans-Neptunian satellite systems are two orders of magnitude larger than the angular momenta of single TNOs of the same masses.

Galimov (2011) and Galimov and Krivtsov (2012) have considered the formation of embryos of the Earth–Moon system in the process of compression of a rarefied condensation. Ipatov (2015a; 2015b; 2015d) has assumed that the parental condensation (with a mass exceeding 0.02 of the Earth mass) that gave birth to these embryos has acquired the greater part of its angular momentum in a collision of two condensations.

Nesvorny et al. (2010) have calculated the compression of preplanetesimals in the trans-Neptunian region and determined the conditions under which this compression ends in formation of binary or triple objects. The authors have assumed that preplanetesimals have acquired their angular momenta in the process of formation from a protoplanetary cloud. Therefore, it was difficult to explain the probable formation of a negative angular momentum of a collapsing condensation and the formed binary object. Nesvorny et al. have noted that the modeling of condensation formation in (Johansen et al., 2007, 2009b) generally ended with direct rotation of rarefied preplanetesimals. Johansen and Lacerda (2010) have demonstrated that solid protoplanets, which had accreted pebble-sized bodies and boulders in a gas medium, acquire prograde rotation. However, the angular momenta of certain known binary trans-Neptunian objects are negative. For example, Sheppard et al. (2012) have examined 17 same-size binary TNOs and found five objects with retrograde rotation. Ipatov (2010) has conjectured that negative angular momenta of certain observed binary objects could be acquired in collisions of preplanetesimals resulting in the formation of preplanetesimals that contracted and gave birth to satellite systems.

The dependences of inclinations of orbits of the secondary components around the primary compo- nents in the known binary TNOs on the distance between the components, on the eccentricity of the orbit of the secondary component around the primary component, on the ratio of diameters of the components, and on the elements of the heliocentric orbit of the binary object were studied in (Ipatov, 2015c; 2017). These dependences were explained using the model of formation of a satellite system in a collision of two rarefied condensations containing dust and/or boulders up to 1 m in diameter. It was assumed that a satellite system was formed in the process of compression of a condensation produced in such a collision. The model of formation of a satellite system in a collision of two condensations is not incompatible with the fact that approximately 40% of binary objects found in the trans-Neptunian belt have a negative angular momen- tum relative to their centers of mass.

Ipatov (2010) has compared the angular momenta of colliding preplanetesimals with the angular momenta of known trans-Neptunian and asteroid satellite systems. After the publication of this study, Nesvorny et al. (2010) have presented the results of modeling of compression of condensations. Their modeling encouraged me to continue the studies of the role of collisions of preplanetesimals in the formation of satellites of small bodies and to obtain new estimates for the stage of rarefied preplanetesimals.

*Observed Fraction of Binary Systems in the Population of Minor Planets*

Noll et al. (2008b) have concluded that the fraction of binary systems in the population of minor planets is 0.3 for cold classical TNOs and 0.1 for all other TNOs. Having examined 477 main-belt asteroids observed prior to May 2011, Pravec et al. (2012) found 45 binary objects (i.e., the fraction of binary objects was 0.094). It is assumed (see, for example, Ipatov, 1987; Levison and Stern, 2001; Gomes, 2003; 2009) that TNOs moving in eccentric orbits ("other TNOs" mentioned above) have formed in the feeding zone of the giant planets (i.e., closer to the Sun than classical TNOs).

Petit and Mousis (2004) have noted that certain celestial bodies, which are now considered single, could have been binary in the past. In addition to classical break-up and ejection of the second component in a high-velocity collision, the authors have investi- gated the possibilities of ejection of the second component from its orbit as a result of a direct collision with a small impactor and of gravitational perturbation of the second component in a close encounter with a massive TNO. According to the obtained estimates, at the current mass of the trans-Neptunian belt, approximately 1/3 of all binary objects could be scattered within a time interval shorter than the age of the Solar System. This fraction may be greater for a greater trans-Neptunian belt mass. The results obtained by Petit and Mousis (2004) suggest that the initial fraction of binary TNOs could be larger than 0.45.

**The following issues are addressed in the present study.** The angular velocities used by Nesvorny et al. (2010) as the initial data for modeling the compression of preplanetesimals are compared with the angular velocity of a The angular velocity of a rarefied preplanetesimal growing via accumulation of small objects is examined. Specifically, the obtained angular velocities are compared with the velocities used by Nesvorny et al. (2010).

Models of preplanetesimal collisions are examined, and mergers of colliding preplanetesimals are discussed.

It is demonstrated that the probability of a collision of a preplanetesimal with other preplanetesimals of a comparable size may be equal to the fraction of TNOs with satellites among all TNOs (if the lifetimes of preplanetesimals given in several studies are considered).

The formation of satellite systems at different distances from the Sun is discussed.

## COMPARISON OF THE ANGULAR MOMENTUM OF TWO COLLIDING PREPLANETESIMALS WITH THE MOMENTUM AT WHICH CONTRACTING PREPLANETESIMALS PRODUCE SATELLITE SYSTEMS

*Model in Which Colliding Preplanetesimals Form a New Preplanetesimal*

The model in which two colliding spherical preplanetesimals form a new spherical preplanetesimal is discussed below. The mass and the angular momentum of the formed preplanetesimal are equal to the sum of masses and the angular momentum of colliding preplanetesimals relative to their common center of mass, respectively. The mass and the angular momentum of a preplanetesimal formed as a result of a merger of two preplanetesimals may in fact be lower than the indicated values, and the formed preplanetesimal may be nonspherical (e.g., a bound pair of preplanetesimals may form).

The angular momentum of two colliding preplanetesimals (with radii $r_1$ and $r_2$ and masses $m_1$ and $m_2$) relative to their center of mass was obtained in (Ipatov, 2010):

$$K_s = k_\Theta \cdot (G \cdot M_S)^{1/2} \cdot (r_1+r_2)^2 \cdot m_1 \cdot m_2 \cdot (m_1+m_2)^{-1} \cdot a^{-3/2}, \qquad (1)$$

where $G$ is the gravitational constant, $M_S$ is the solar mass, and the difference between semimajor axes $a$ of preplanetesimals is $\Theta(r_1+r_2)$. $k_\Theta$ is discussed in the next paragraph. It was assumed in the derivation of (1) that preplanetesimals moved in circular heliocentric orbits prior to the collision. If preplanetesimals had angular momenta before the collision, these momenta should be added to the above $K_S$ value.

At $r_a=(r_1+r_2)/a<<\Theta$ and $r_a<<1$, one may obtain $k_\Theta \approx (1-1.5\cdot\Theta^2)$. The average value of $|k_\Theta|$ is 0.6. Angular momentum $K_S$ is positive at $0<\Theta<(2/3)^{1/2}\approx 0.8165$ and negative at $0.8165<\Theta<1$. The minimum $k_\Theta$ value is $-0.5$. In the case of a uniform $\Theta$ distribution, the probability to acquire prograde rotation in a single collision is ~0.2; if multiple collisions are considered, the ratio of the sum of positive $K_S$ values to the sum of absolute values of negative $K_S$ values is 9.4. The fraction of collisions with a negative angular momentum may differ from 20% if the mutual gravitational influence of preplanetesimals prior to their collision and/or eccentric heliocentric orbits are considered.

Ratio $r_K$ of the angular momentum of the entire satellite system to the angular momentum of the primary component relative to its center of mass was as high as 165 for the known binary TNOs considered in (Ipatov, 2010). The ratio of the maximum angular momentum, which may be acquired in a collision of two preplanetesimals with radii equal to their Hill radii, to the angular momentum of the formed primary component of a satellite system may be even higher than the above $r_K$ value.

It is assumed in the model considered below that colliding preplanetesimals of one and the same density merge and form a spherical preplanetesimal with radius $r=(r_1^3+r_2^3)^{1/3}$. Angular velocity $\omega$ of the formed preplanetesimal is $K_S/J_S$, where $J_S=0.4\chi(m_1+m_2)r^2$ is the moment of inertia of the formed preplanetesimal with radius $r$, and $\chi=1$ for a homogeneous sphere considered by Nesvorny et al. (2010). Ipatov (2010) has obtained the following:

$$\omega=K_s/J_s=2.5\cdot k_\Theta \cdot \chi^{-1} \cdot (r_1+r_2)^2 \cdot r^{-2} \cdot m_1 \cdot m_2 \cdot (m_1+m_2)^{-2}\Omega,$$

where $\Omega=(G\cdot M_S)^{1/2}a^{-3/2}$ is the angular velocity of motion of a preplanetesimal around the Sun. If radius $r_c$ of a of a spherical preplanetesimal, which was formed in a collision of homogeneous spherical preplanetesimals with radii $r_1$ and $r_2$, is $k_{rc}r$ after compression at a certain time after the collision, then the angular velocity of the compressed preplanetesimal is $\omega_c=\omega\cdot k_{rc}^{-2}$ at $\chi=1$.

*Angular Velocity of a Preplanetesimal Needed to Form a Satellite System*

The compression of preplanetesimals in the trans-Neptunian belt was modeled by Nesvorny et al. (2010) for the following initial angular velocities of preplanetesimals: $\omega_o = k_\omega \Omega_o$, where

$\Omega_O = (Gm)^{1/2} r^{-3/2}$ is the angular velocity of motion in a circular orbit with radius $r$ around a gravitating center of mass $m$. The following values of $k_\omega$ were used: 0.5, 0.75, 1, and 1.25. In most calculations, $r = 0.6 r_H$, where $r_H$ is the Hill sphere radius for mass $m$. Nesvorny et al. (2010) have also considered $r = 0.4 r_H$ and $r = 0.8 r_H$. Note that $\Omega_O/\Omega = 3^{1/2}(r_H/r)^{3/2} \approx 1.73 (r_H/r)^{3/2}$ (e.g., $\Omega_O \approx 1.73\Omega$ at $r = r_H$).

In the case when $r_1 = r_2$, $r^3 = 2 r_1^3$, $m_1 = m_2 = m/2$, and $\chi = 1$, which is considered below, we obtain $\omega = 1.25 \times 2^{1/3} k_\Theta \Omega \approx 1.575 k_\Theta \Omega$. For example, $\omega \approx 0.945\Omega$ at $k_\Theta = 0.6$. Using the above formulas for $\omega$ and $\omega_o$ and if the radii of preplanetesimals are equal to the radii of their Hill spheres, we arrive at the conclusion that $\omega = \omega_o$ at $k_\omega = 1.25 \times 2^{1/3} \times 3^{-1/2} k_\Theta \chi^{-1} \approx 0.909 k_\Theta \chi^{-1}$. Since the value of $k_\Theta$ may vary through to 1, this ratio demonstrates that values of $\omega = \omega_o$ corresponding to $k_\omega$ up to 0.909 may be obtained in collisions of homogeneous ($\chi = 1$) preplanetesimals.

In the case of a collision of two preplanetesimals with radii equal to the radii of their Hill spheres and subse- quent compression of the produced preplanetesimal to radius $r_c$ (at $\chi = 1$), the angular velocity of the compressed preplanetesimal is $\omega_{rc} = \omega_H (r_H/r_c)^2$, where $\omega_H \approx 1.575 k_\Theta \Omega \approx 0.909 k_\Theta \Omega_{oH}$ and $\Omega_{oH} = (G \cdot m)^{1/2} r_H^{-3/2}$. Since the initial angular velocity of the preplanetesimal that starts contracting is $\omega_o = k_\omega \Omega_o = k_\omega (G \cdot m)^{1/2} r_c^{-3/2}$ at $r = r_c$, then $\omega_{rc}/\omega_o \approx 0.909 (k_\Theta/k_\omega)(r_H/r_c)^{1/2}$. If we analyze the collision of two preplanetesimals with Hill radii, the angular velocity of the formed preplanetesimal after its compression to radius $rc = 0.6 r H$, which was considered in Nesvorny et al. (2010), corresponds to $k_\omega$ as high as $0.909 \times 0.6^{-1/2} \approx 1.17$. In the calculations of Nesvorny et al. (2010), binaries or tripples were obtained only at $k_\omega = 0.5$ or 0.75. As noted above, such $k_\omega$ values may be obtained in collisions of preplanetesimals at $k_\Theta = 1$ ($k_\omega = 0.5$ may also be obtained at $|k_\Theta| = 0.5$). No satellite systems were produced in the calculations of Nesvorny et al. (2010) at $k_\omega = 1$ or 1.25. Thus, one may conclude that *the initial angular velocities of preplanetesimals corresponding to the formation of satellite systems in the calculations of Nesvorny et al. (2010) could be acquired in collisions of preplanetesimals that produced parental preplanetesimals.*

*Discussion of Other Models of Collisions of Preplanetesimals*

In the model considered above, the mass of the preplanetesimal formed in a collision is equal to the sum of masses of colliding preplanetesimals. In actual collisions, the mass of the formed preplanetesimal may be lower than the sum of masses of colliding preplanetesimals, or the greater part of mass may remain in the collided preplanetesimals that do not form a new preplanetesimal. If preplanetesimals are assumed to move in unperturbed circular heliocentric orbits prior to their collision, the collision is more grazing at higher Θ values; therefore, the ratio of the mass of the formed preplanetesimal to $m_1 + m_2$ may be lower than that at a lower Θ. In the case of initially circular helio- centric orbits, the relative motion of the centers of mass of encountering preplanetesimals may become rather complex if the mutual gravitational influence of mass of preplanetesimals is considered. If this is the case, collisions corresponding to negative angular momenta may occur at such a difference between the semimajor axes of orbits that exceeds the Hill sphere radius for mass $m_1 + m_2$ (Eneev and Kozlov, 1981; 2016). The above factors may alter the estimate of the fraction of binary systems with a negative angular momentum obtained using the model from the previous subsection.

It is our belief that a solid object with two contacting components may be produced if two compression centers are present in a preplanetesimal, which had formed as a result of merger of two collided rarefied condensations, at such values of the angular momen- tum of the preplanetesimal that are somewhat lower than the ones needed to produce binary objects. This mechanism of formation of a pair of contacting bodies may account for the fact that certain observed asteroids and comets (e.g., Itokava and 67P/Churyumov– Gerasimenko) have two parts and the shape of a dumbbell. This mechanism of production of dumbbell-shaped objects requires the masses of the smallest formed condensations to be equal to the masses of such asteroids and comets.

*Contribution of the Initial Angular Momenta of Collided Preplanetesimals to the Angular Momentum of the Formed Condensation*

The collided condensations could have nonzero angular momenta prior to the collision. Therefore, the angular velocity of the formed condensation could exceed the value of $0.9\Omega_o$ obtained above. According to Safronov (1972), initial angular velocity $\omega_{of}$ of a rarefied condensation was $0.2\Omega$ for a spherical condensation and $0.25\Omega$ for a flat disk, where $\Omega$ is the angular velocity of motion of a condensation around the Sun. It was assumed in the derivation of these estimates that the average angular momentum of a condensation is close to the angular momentum (relative to the condensation center) of the region from which the condensation was formed, and each element of volume of this region is in an unperturbed Keplerian circular motion around the Sun. The initial angular velocity of the condensation is positive. It follows from the comparison of $0.2\Omega$ with the values of $\omega_o$ from (Nesvorny et al., 2010) that this angular velocity is not sufficient to form satellites. If two identical *homogeneous* spherical condensations with initial angular velocity $\omega_{of}$ collided with no additional relative angular momentum, then the angular velocity of the spherical condensation formed in this collision is $\omega_2 = 2^{-2/3}\omega_{of}$; for example, $\omega_2 = 0.126\Omega$ at $\omega_{of} = 0.2\Omega$. It was demonstrated above that the angular velocity of the parental condensation formed in a collision of two identical condensations is $\omega \approx 1.575 k_\Theta \cdot \Omega$ (at any $r/r_H$). At $k_\Theta = 0.6$ we have $\omega/\omega_2 = 0.945/0.126 = 7.5$, at $k_\Theta = 1$, $\omega/\omega_2 \approx 12.5$. These $\omega/\omega_2$ values demonstrate that the contribution of $\omega$ to the resulting angular velocity $\omega + \omega_2$, which is produced by a typical collision of identical homogeneous condensations, is several times larger than angular velocity $\omega_2$ associated with the initial condensation rotation.

The above estimates were obtained under the assumption that the sizes of colliding preplanetesimals are the same as the sizes of preplanetesimals at the time when they acquired their initial rotation. The contribution of the initial angular momenta of collided preplanetesimals to angular momentum $K_s$ of the formed parental preplanetesimal, which contracted and formed a satellite system or a single object, may be greater if the preplanetesimals did decrease in size prior to their collision. Let us consider the collision of two *identical homogeneous* preplanetesimals with masses $m_1$ and radii equal to $k_{col}r_H$ (Hill sphere radius $r_H = a \cdot [m_1/(3 \cdot M_S)]^{1/3}$). We assume that each of these preplanetesimals was formed with an initial radius of $k_{in}r_H$ and an angular velocity of $0.2\Omega$ (i.e., with an angular momentum of $0.2\Omega J_s$, where $\Omega = G^{1/2} \cdot M_S^{1/2} \cdot a^{-3/2}$.). The moment of inertia of the initial preplanetesimal is $J_s = 0.4\chi m_1(k_{in}r_H)^2$ ($\chi$ characterizes the distribution of matter within the initial preplanetesimals; $\chi = 1$ for homogeneous spheres).

The angular momentum of a spherical preplanetesimal formed in such a collision is
$K_s = (2k_\Theta \cdot k_{col}^2 + 0.16 \cdot \chi \cdot k_{in}^2) \cdot r_H^2 \cdot a^{-3/2} \cdot m_1 \cdot G^{1/2} \cdot M_S^{1/2} \approx$
$\approx (0.96 k_\Theta \cdot k_{col}^2 + 0.077 \cdot \chi \cdot k_{in}^2) a^{1/2} \cdot m_1^{5/3} \cdot G^{1/2} \cdot M_S^{-1/6}$. Thus, the initial rotation is more important than the collision if $\chi = 1$ at $k_{in}/k_{col} > 2.7$ and $k_\Theta = 0.6$ (or at $k_{in}/k_{col} > 3.5$ and $k_\Theta = 1$). This suggests that *collisions produced the dominant contribution to angular momentum $K_s$ only when the sizes of homogeneous preplanetesimals did not differ significantly (by not more than a factor of 3) from the initial sizes. The sizes differed by a factor of no more than $3\chi^{-1/2}$ in the case of inhomogeneous (denser at the core) preplanetesimals.*

Let us consider the merger of two preplanetesimals with masses $k_m m$ and $(1 - k_m)m$ ($0 < k_m < 1$) and an initial angular velocity of $0.2\Omega$. We assume that $\chi$ is the same for all the considered preplanetesimals (including the formed preplanetesimal). If the densities of both preplanetesimals at the moment of their formation were the same and equal to $\rho$, the component of angular momentum $K_s$ of the produced preplanetesimal (with radius $r$) associated with the initial rotation $K_{si} = 0.2\Omega(0.4\chi \cdot m \cdot r_{in}^2)[(1-k_m)^{5/3} + k_m^{5/3}]$, where $r_{in}$ is the radius of a preplanetesimal with mass $m$ and den- sity $\rho$. The collision-related component of $K_s$ of the formed preplanetesimal with radius $r_{col}$ is $K_{sc} = k_\Theta \cdot \Omega \cdot m \cdot r_{col}^2 \cdot k_m(1-k_m) \cdot [(1-k_m)^{1/3} + k_m^{1/3}]^2$. If the preplanetesimals had contracted prior to the collision, then $r_{col}$ is smaller than $r_{in}$. At $k_\Theta = \chi = 1$ and $r_{col} = r_{in}$, ratio $K_{sc}/K_{si}$ is 12.5, 3, and 0.8 at $k_m = 2^{-1}$, $9^{-1}$, and $28^{-1}$ (i.e., at ratio $k_r = 1$, 2, and 3 of the radii of colliding preplanetesimals), respectively. This means that *if ratio $k_r$ of the radii of colliding homogeneous preplanetesimals is larger than three in the model under consideration, the contribution of the initial rotation to angular momentum $K_s$ of the formed preplanetesimal is more significant than that of the collision.* If $\chi$ is below 1 (i.e., preplanetesimals are denser at the cores), the contribution of the initial rotation to angular momentum $K_s$ of the formed preplanetesimal is lower than that at $\chi = 1$. *In the case of $\chi < 1$, the contribution of the collision to the*

*angular momentum of the formed preplanetesimal may be larger than the contribution of the initial rotation even at* $k_r > 3$.

The term "close preplanetesimal sizes" is used below in the case when the collision of considered preplanetesimals produces the dominant contribution to the angular momentum of the parental preplanetesimal. For example, if the ratio of diameters of collided homogeneous preplanetesimals, which moved in almost circular heliocentric orbits prior to the collision, does not exceed three. In certain collisions, the mass of the formed preplanetesimal may be smaller than the sum of masses of the collided preplanetesimals, and a fraction of the angular momentum of the collided preplanetesimals may be carried away by matter that did not become a part of the new preplanetesimal.

The angular momentum of the initial preplanetesimals was positive, and the fact that approximately 40% of the observed trans-Neptunian satellite systems have a negative angular momentum (Ipatov, 2017) may, in our opinion, be attributed to preplanetesimal collisions that resulted in the formation of parental preplanetesimals. Therefore, the above estimates suggest that *the radii of the majority of collided homogeneous preplanetesimals (condensations), which formed parental preplanetesimals for trans-Neptunian satellite systems, decreased by a factor of no more than three in the interval from the moment of formation to the moment of collision, and the radii of two collided preplanetesimals differed by a factor of no more than three. These differences could be more significant in the case of inhomogeneous preplanetesimals.*

## ANGULAR MOMENTUM OF A PREPLANETESIMAL FORMED VIA ACCUMULATION OF SMALL OBJECTS

Johansen and Lacerda (2010) have obtained direct axial rotation in the hydrodynamic modeling of accretion of pebble- and boulder-sized objects onto protoplanets with a radius of several hundred kilometers in a gas medium. To explain the observed rotation of the largest asteroids and TNOs, the authors have assumed that protoplanets had acquired a considerable (10–50%) fraction of their mass via accumulation of pebble- and boulder-sized objects in the gas phase of the solar nebula. The angular momentum of a rarefied preplanetesimal formed via accumulation of small objects is analyzed below. The formulas and certain results of examination of this angular momentum were presented in the fourth section of the fifth chapter of (Ipatov, 2000). Several formulas and brief results may also be found in (Ipatov, 2010). More detailed data were presented in (Ipatov, 1981a; 1981b). These studies were focused on the axial rotation of planets. The same formulas are applied below to the process of formation of binary objects to determine whether it is possible for binary small bodies to acquire a significant fraction of their angular momentum in a model where the parental rarefied preplanetesimal accumulates small objects. Johansen and Lacerda (2010), Lambrechts and Johansen (2012), and Johansen et al. (2015b) have considered the growth of planetesimals in the process of accumulation of objects less than 1 m in size. In the models considered below, objects falling onto a preplanetesimal or a planetesimal may be larger if their mass is still much lower than that of a preplanetesimal (planetesimal).

### *Positive and Negative Increments of the Angular Momentum of a Preplanetesimal*

When small objects fall onto a preplanetesimal (planetesimal), the angular momentum (AM) acquired by it is proportional to difference $\Delta K = K^+ - K^-$ between positive $K^+$ and negative $K^-$ AM increment fractions of this preplanetesimal (planetesimal) ($K^+ + K^- = 1$). For example, $K^-$ is equal to the ratio of the sum of moduli of negative AM increments to the sum of moduli of all (positive and negative) AM increments. The values of $\Delta K$ for different eccentricities and semimajor axes of heliocentric orbits of objects encountering with a preplanetesimal at a distance smaller than $r_s$ (the radius of the considered sphere) of a preplanetesimal were presented in (Ipatov, 1981a; 1981b) and (in brief) in (Ipatov, 2000). Spheres of action ($r_s=a \cdot m_r^{2/5}$, where $m_r$ is the ratio of the preplanetesimal mass to the solar mass) and spheres of influence ($r_s=1.15a \cdot m_r^{1/3}$) were considered. Hundreds of thousands of close approaches of objects to a preplanetesimal were modeled with different masses and elements of heliocentric orbits of objects. The motion of objects was modeled using the spheres method: objects and a preplanetesimal moved in unperturbed heliocentric orbits outside the sphere, and their relative motion inside the sphere was simulated by solving a two-body problem. The values of $\Delta K$ and the probability of a collision between objects and a preplanetesimal were calculated for various preplanetesimal diameters that did not exceed the diameters of the considered spheres. The $\Delta K$ values obtained varied from less than –0.4 to almost 1.

The results of computer simulation presented in (Ipatov, 1981a; 1981b) and (in brief) in (Ipatov, 2000) demonstrate that $\Delta K \approx 0.9$ if a preplanetesimal with a size of the considered sphere and small objects falling onto it moved in circular heliocentric orbits prior to the collision. This estimate agrees with the theoretical ratio of $K^+/K \approx 9.4$ that was obtained in (Ipatov, 2010) for circular heliocentric orbits of a preplanetesimal and small objects distributed uniformly over the semimajor axes. The values of $\Delta K$ obtained for a preplanetesimal with mass m and a radius close to the radius of its sphere of influence or sphere of action at eccentricities $e<5(m/M_S)^{1/3}$ of heliocentric orbits of objects were generally larger than 0.6. The values of $\Delta K$ for more eccentric orbits may be much lower than those for circular orbits and may become negative. The $\Delta K$ values for different semimajor axes of heliocentric orbits of objects may differ even if the eccentricities of these orbits remain the same.

*Growth of a Preplanetesimal of the Same Size as Its Hill Sphere*

Let us consider the collision of two spherical pre- planetesimals with radii $r_1$ and $r_2$ that moved in the same plane in circular heliocentric orbits with difference between their semimajor axes a prior to the collision equal to $\Theta(r_1 + r_2)$. Their mutual gravitational influence is neglected. The collision velocity of these preplanetesimals is then written as

$$v_{col}=v_c(r_1+r_2)a^{-1}(1-0.75\Theta^2)^{1/2}, \qquad (2)$$

and its tangential component is

$$v_\tau=v_c(r_1+r_2)a^{-1}(1-1.5\Theta^2)=v_c(r_1+r_2)a^{-1}k_\Theta, \qquad (3)$$

where $v_c=(G \cdot M_S/a)^{1/2}$ is the velocity of motion of a preplanetesimal around the Sun (Ipatov, 2010). The averaged (over all $\Theta$ values) value of $|v_\tau|$ is $0.6v_c(r_1+r_2)a^{-1}$.

If radius r of a growing preplanetesimal is $k_H r_H$ ($k_H$ is a constant, and $r_H$ is the Hill radius of this preplanetesimal) and $|v_\tau|=0.6v_c \cdot r \cdot a^{-1}$, then the angular momentum $K_s$ of a preplanetesimal with mass $m_f$, which grew by accumulating small objects, is

$$K_s \approx 0.173 k_H^2 G^{1/2} a^{1/2} m_f^{5/3} M_S^{-1/6} \Delta K. \qquad (4)$$

Formula (4) was obtained by integrating the angular momentum increment over mass m from 0 to $m_f$. It was taken into account that the angular momentum increment is $dK_s=r \cdot v_\tau \cdot dm$, $dm=4\pi \cdot \rho \cdot r^2 dr$ and $m=4\pi \cdot \rho \cdot r^3/3$ (density $\rho$ of a growing preplanetesimal was assumed to be constant). If the growth of the preplanetesimal mass from $m_o$ to $m_f$ is analyzed, then $m_f^{5/3}$ in the above formula should be replaced with $m_f^{5/3}-m_o^{5/3}$.

Since $K_s=J_s\omega=2\pi J_s/T_s$, $J_s=0.4 \cdot \chi \cdot m_f(k_H \cdot r_{Hf})^2$ and $r_{Hf} \approx a(m_f/3M_s)^{1/3}$, it follows from (4) that the axial rotation period of a preplanetesimal growing via collisions with smaller objects is $T_s \approx 7 \cdot \chi \cdot a^{3/2} \cdot (G \cdot M_S)^{-1/2} \cdot \Delta K^{-1}$, if $|v_\tau|=0.6v_c \cdot r \cdot a^{-1}$. In the considered model, where the radius of a growing preplanetesimal equals $k_H \cdot r_H$, $T_s$ does not depend on $k_H$. The angular velocity of this preplanetesimal after its compression to radius $r_c=k_{rH} r_H$ is $k_{rH}^{-2}$ times higher than the angular velocity before the compression: $\omega_c \approx 0.9 \cdot \chi^{-1} \cdot a^{-3/2} \cdot (G \cdot M_S)^{1/2} \cdot \Delta K \cdot k_{rH}^{-2}$. If we assume that $\omega_c=\omega_o=k_\omega \Omega_o=k_\omega(G \cdot m)^{1/2} r_c^{-3/2}$ and $r_c \approx k_{rH} a(m/3M_S)^{1/3}$, then $k_\omega \approx 0.9 \cdot \chi^{-1} \cdot 3^{-1/2} \Delta K \cdot k_{rH}^{-2}$. At $\Delta K=0.9$ this $\Delta K$ value was obtained with almost circular heliocentric orbits of preplanetesimals with their radii close to the radii of their Hill spheres) and $\chi = 1$, we obtain $k_\omega \approx 0.47 k_{rH}^{-2}$, for example, $k_\omega \approx 1.3$ at $k_{rH}=0.6$ and $k_\omega \approx 0.73$ for $k_{rH}=0.8$. If we consider the compression of a sphere with radius $k_H \cdot r_H$ by a factor of $k_{rH}^{-1}$, then the radius of the formed preplanetesimal is $0.6r_H$ at $k_{rH}=0.8$ and $k_H=0.75$ (as in the calculations of Nesvorny et al. (2010)), and $k_\omega \approx 0.73$. The above estimates suggest that a preplanetesimal formed via accumulation of small objects, which moved in almost circular heliocentric orbits prior to the collision with a condensation, may acquire such angular velocity values (corresponding to $k_\omega = 0.5$ and 0.75) that were used by Nesvorny et al. (2010) as the initial data for calculating the compression of a preplanetesimal resulting in the formation of a satellite system.

The above estimates demonstrate that a parental preplanetesimal, which has a radius close to the Hill radius and has acquired a considerable fraction of its mass in multiple collisions with small objects moving in almost circular heliocentric orbits, may acquire such angular momentum values that induce the formation of a solid TNO with satellites (instead of a single body). Approximately 40% of binary TNOs have negative angular momenta (see (Ipatov, 2017)), while the resulting angular momentum of a preplanetesimal after accretion of many small objects from weakly eccentric orbits is positive. Therefore, one may conclude that in most cases, the greater part of the angular momentum of binary TNOs was acquired not in the process of accumulation of small objects, but in a collision of two preplanetesimals. However, a certain fraction of the angular momentum of binary TNOs might be delivered to their parental preplanetesimals by small objects. The masses of TNOs could vary over the course of evolution of the Solar System due to collisions with other objects. In principle, if the eccentricities of orbits of objects colliding with a preplanetesimal are sufficiently large, retrograde rotation of this preplanetesimal is possible; however, almost all preplanetesimals of the same size (or smaller) in the same feeding zone should then have retrograde rotation. In our opinion, such eccentricities are also not likely to be reached at the stage of rarefied condensations.

*Growth of a Preplanetesimal Much Smaller in Size Than Its Hill Sphere*

The above estimates correspond to the case when small objects move in unperturbed heliocentric orbits prior to their collision with a preplanetesimal. Specifically, the radii of preplanetesimals equal to the radii of their Hill spheres were considered. Taking the motion of objects inside the Hill sphere into account, Ipatov (2000) has obtained the following average tangential component of the collision velocity for a preplanetesimal (planetesimal) that is smaller than its Hill sphere and grew in collisions with smaller objects: $v_\tau = 0.6 v_{par}$, where $v_{par}$ is the parabolic velocity at the preplanetesimal surface. This value was obtained in simulations (by the spheres method) of motion of a large number (hundreds of thousands in each run) of encounters of an object to a preplanetesimal (with different orbits and masses in different runs).

The following formula may be obtained (Ipatov, 2000) for a preplanetesimal with a radius being much smaller than the radius of its Hill sphere at $v_\tau = \alpha v_{par}$ in the model of preplanetesimal growth via accumulation of small objects:

$$K_s \approx 0.67 \alpha \cdot G^{1/2} \rho^{-1/6} m_f^{5/3} \Delta K. \tag{5}$$

Density $\rho$ of a preplanetesimal with its mass $m$ growing from 0 to $m_f$ was assumed to be constant in the derivation of this formula. As formula (4) for $K_s$ from the previous subsection, formula (5) was obtained by integrating the angular momentum increment over mass $m$ from 0 to $m_f$. It was also considered that $v_{par} = (2Gm/r)^{1/2}$.

If radius $r_p$ of the formed preplanetesimal is equal to $k_H r_H$ ($r_H$ is the Hill radius of this preplanetesimal) and taking relations $\rho^{-1/6} = (4\pi/3m_f)^{1/6}(k_H r_H)^{1/2}$ and $r_H \approx a(m_f/3M_S)^{1/3}$ into account, one obtains the following from (5):

$$K_s \approx \alpha \cdot 0.71 k_H^{1/2} G^{1/2} a^{1/2} m_f^{5/3} M_S^{-1/6} \Delta K. \tag{6}$$

If the mass grows from $m_{fo}$ to $m_f$, $m_f$ in formulas (5)–(6) is replaced with $(m^{5/3} - m^{5/3})$. Note that in the model with $|v_\tau| = \alpha \cdot v_{par}$, the dependence of $K_s$ on $k_H$ ($k_H$ instead of $k_H$ in formula (4)) is weaker than that in the model where preplanetesimals move in unperturbed circular heliocentric orbits prior to the collision and $v_\tau = 0.6 v_c$ ra. The actual dependence of $K_s$ on $k_H$ is somewhere between the estimates for these two models. In both models, $K_s$ is lower at lower $k_H$ values.

If we rely solely on the calculations of Nesvorny et al. (2010), in which the radius of a preplanetesimal exceeded 40% of its Hill radius, then it cannot be claimed that a satellite system could form because of compression of a parental preplanetesimal (formed via accumulation of small objects or in a collision of two preplanetesimals) with its radius being sufficiently small relative to its Hill radius. No other calculations of compression of TNO-producing preplanetesimals have been performed. Galimov and Krivtsov (2012) have calculated the compression of condensations with their masses equal to that of the Earth–Moon system and their radii being 5.5 times larger than the radius of a body with a density of the Earth and a mass equal to the mass of this system (i.e., the condensation radius was approximately 40 times shorter than the radius of the corresponding Hill sphere). In two-dimensional calculations, the authors have obtained the formation of a binary system at $0.64 \leq \omega_o/\Omega_o \leq 1.1$ ($\omega_r \approx 1.535\Omega_o$ was considered), and several satellites formed at $\omega_o/\Omega_o > 1.1$. In certain cases, satellite systems formed at $1 \leq \omega_o/\Omega_o \leq 1.5$ in the three-dimensional model. These $\omega_o/\Omega_o$ values are approximately two times higher than the ones (0.5 and 0.75) corresponding to the formation of satellite systems in the calculations of Nesvorny et al. (2010) for much lower preplanetesimal masses and much higher ratios of the preplanetesimal radius to the Hill radius. The calculations in (Galimov and Krivtsov, 2012) suggest that satellite systems of TNOs may sometimes form at lower (compared to the ones obtained in (Nesvorny et al., 2010)) ratios of the preplanetesimal radius to the Hill radius and lower angular momenta, but not at lower $\omega_o/\Omega_o$ ratios. It would be instructive to analyze the compression of preplanetesimals with a set of initial data that is much more extensive than the ones used in (Nesvorny et al., 2010; Galimov and Krivtsov, 2012).

Using formulas (4)–(5) and relations $T_s = 2\pi J_s/K_s$, $J_s = 0.4\chi \cdot m_f \cdot r_p^2$ and $T_{s\rho} = T_s \cdot k_{r\rho}^{-2}$ (where $k_{r\rho}$ is the ratio of radius $r_p$ of a preplanetesimal to the radius of a solid planetesimal that was formed via compression of this rarefied preplanetesimal to density $\rho_s$), one may obtain the formulas for period $T_{s\rho}$ of axial rotation of a planetesimal that grew by accumulating small objects:

$$T_{s\rho} \approx 7\chi \cdot a^{3/2}(G \cdot M_S)^{-1/2} k_{r\rho}^{-2} \cdot \Delta K^{-1} \text{ at } |v_\tau| = 0.6 v_c \cdot r \cdot a^{-1} \text{ and}$$
$$T_{s\rho} \approx 1.45\chi(\rho_s G \cdot k_{r\rho})^{-1/2}(\alpha \cdot \Delta K)^{-1} \text{ at } |v_\tau| = \alpha \cdot v_{par} . \qquad (7)$$

At $\rho_s = 2$ g cm$^{-3}$, $k_{r\rho} = 1$ (bodies fall onto a solid planetesimal), $\chi = 1$, and $\alpha = 0.6$, $T_{s\rho} \approx 1.84/\Delta K$ h is obtained using (7). It can be seen that in the case of exceeded 40% of its Hill radius, then it cannot be claimed that a satellite system could form because of solid-body accumulation, the periods of axial rotation of small bodies vary with $\Delta K$.

*The considered model of growth of the angular momentum of a preplanetesimal (planetesimal) that is much smaller than its Hill sphere may be used to study the evolution of axial rotation of small bodies and the growth of angular momenta of embryos of binary objects.*

One of the possible values of the axial rotation period of a small body is close to 6 h. For example, this period for the (38628) Huya TNO is 6.75 h. At $T_{s\rho} = 6$ h and $|v_\tau| = 0.6 v_{par}$ it follows from (7) that $\Delta K = K^+ - K^-$ is written as

$$\Delta K \approx 0.16 \cdot k_{r\rho E}^{-1/2}, \qquad (8)$$

where $K^+$ is the ratio of the sum of modules of positive angular momentum (AM) increments to the sum of modules of all (positive and negative) AM increments, $K^-$ is the ratio of the sum of moduli of negative AM increments to the sum of moduli of all AM increments, and $k_{r\rho E}$ is the ratio of the preplanetesimal radius to the radius of a solid planetesimal formed by compression of the preplanetesimal to density $\rho_S$ equal to the density of the Earth. If a preplanetesimal is compressed to a density of 2 g cm$^{-3}$, the corresponding formula is $\Delta K \approx 0.27 \cdot k_{r\rho 2}^{-1/2}$, where $k_{r\rho 2}$ is the coefficient for a density of 2 g cm$^{-3}$. Since $k_{r\rho E} \geq 1$, formula (8) holds true at $\Delta K \leq 0.16$ (equality is obtained when bodies fall onto a solid planetesimal). The higher $k_{r\rho E}$ is, the lower is the maximum allowed $\Delta K$ value. For example, $\Delta K \approx 0.01$ at $k_{r\rho E}$ (the ratio of the Hill radius of the Earth to the radius of the Earth) being approximately equal to 235. At one and the same ratio of the preplanetesimal radius to the Hill radius, the values of $\Delta K$ in formula (8) at 36 AU from the Sun are six times lower than those at 1 AU. It can be seen from (7) that the values of $\Delta K$ go down as $T_{S\rho}$ increases.

A considerable fraction of angular momentum of an uncompressed preplanetesimal may be carried away by the material that escapes the preplanetesimal in the process of its compression. The angular momentum in (4)–(6) is inversely proportional to $\Delta K$. The results of calculations presented in (Ipatov, 1981a; 1981b) and briefly mentioned in (Ipatov, 2000) demonstrate that $\Delta K$ tends to decrease with an increase in eccentricities and semimajor axes of heliocentric orbits of objects falling onto a preplanetesimal (planetesimal). If the reduction in the angular momentum of a contracting preplanetesimal is taken into account, larger $\Delta K$ values (compared to the ones from the previous paragraph) and lower eccentricities of heliocentric orbits of accreted bodies may correspond to preplanetesimals that, when compressed, form solid bodies with rotation periods of several hours. However, if a preplanetesimal with a radius of at least 0.1 of its Hill sphere radius acquires a considerable fraction of its mass by accumulating small objects with almost circular orbits ($\Delta K$ is then close to 1) in the trans-Neptunian region, then the corresponding axial rotation periods of solid planetesimals formed after compression (calculated using the formula (7)) are shorter than the minimum possible axial rotation period of 2.2 h (Rozitis et al., 2014) even when 90% of the angular momentum are lost in the process of compression. In the model with $|v_\tau|=0.6v_c \cdot r \cdot a^{-1}$, the angular momentum of a preplanetesimal with a radius close to its Hill radius, which was formed either in a collision of two identical planetesimals or via accumulation of small objects, is more than 100 times higher than the typical angular momenta of the observed TNOs of the same mass. It was mentioned in the last subsection of the first section that the angular momentum of an initial homogeneous preplanetesimal is not more than an order of magnitude lower than the angular momentum at a collision of two identical preplanetesimals. Therefore, if the radii of initial homogeneous preplanetesimals are close to their Hill radii, the angular momenta of these preplanetesimals are also considerably larger than the angular momenta of observed single TNOs of the same mass. Thus, *when a preplanetesimal with a size of its Hill sphere is compressed to the density of solid bodies without the formation of satellites, most of its angular momentum is carried away by the escaping material.*

*Relative Growth of Solid Planetesimals and Their Satellites*

The formed solid planetesimals and their satellites could grow by accumulating other objects. Owing to runaway accretion, the growth rate of large planetesimals is higher than that of small planetesimals and satellites. The effective radius (capture radius) $r_{ef}$ of a planetesimal (preplanetesimal) with radius $r$ is the impact parameter at which the trajectory of an object falling onto this planetesimal (preplanetesimal) touches its surface. The following relation holds true: $r_{ef} \approx r[1+(v_{par}/v_{sH})^2]^{1/2}$, where $v_{par}=(2G \cdot m/r)^{1/2}$ is the parabolic velocity at the surface of a planetesimal with mass $m$, and $v_{sH}$ is the velocity of a body, which enters the Hill sphere of a planetesimal, relative to this planetesimal. At high eccentricities of heliocentric orbits, $r_{ef}$ is close to $r$.

If $(v_{par}/v_{sH})^2 \gg 1$, then $r_{ef} \approx (2G \cdot m/r)^{1/2}/v_{sH}$. Since $m$ is proportional to $r^3$, $r_{ef}$ is proportional to $r^2$ in this case. This proportionality is retained in the case of almost circular heliocentric orbits of planetesimals with radius $r$ and smaller objects falling onto them (Safronov, 1972, p. 132).

If effective radius $r_{ef}$ of a planetesimal is proportional to $r^2$, the probability that a body travels at a minimum distance of $r_e$ from the center of mass of a planetesimal increases at lower $r_e$ values. This is attributed to the fact that minimum distance $r_e$ may be attained at an impact parameter that is much larger than $r_e$. Therefore, owing to the influence of the gravitational field of a planetesimal, the probability of collision of a body with a close satellite of a planetesimal may be higher than the probability of collision with a more distant satellite of the same size.

The increment of the angular momentum of a planetesimal at the stage of solid-body accretion may be positive and may exceed (in modulus) the angular momentum of this planetesimal at the moment of its formation from a rarefied condensation. The angular momentum of a planetesimal may then become positive even if its angular momentum and the angular momentum of the entire planetesimal–satellite system were initially negative.

Likewise, the angular momentum of a planetesimal may become negative even if its angular momentum and the angular momentum of the entire planetesimal–satellite system were initially positive. *It follows that the angular momentum of a planetesimal relative to its center of mass and the angular momentum of its satellite relative to the planetesimal may have different signs.*

The relative growth of mass $m_p$ of a planetesimal and mass $m_s$ of its satellite may be examined using the formulas given below. If $r_{ef}$ is proportional to $r$, relation $dm_s/m_s = k_1(m_s/m_p)^{2/3} dm_p/m_p$, where $k_1 = k_d^{2/3}$, $k_d$ is the ratio of densities of the planetesimal and the satellite) is valid. Integrating this relation, one may obtain $m_s/m_{so} = [k_o^{-2/3} + k_1(k_m^{-2/3} - 1)]^{-3/2}/k_o$, where $k_o = m_{so}/m_{po} \leq 1$, $k_m = m_p/m_{po} \geq 1$, and $m_{so}$ and $m_{po}$ are the initial values of $m_s$ and $m_p$, respectively. If $r_{ef}$ is proportional to $r^2$, then $m_s/m_{so} = [k_o^{-4/3} + k_2(k_m^{-4/3} - 1)]^{-3/4}/k_o$, where $k_2 = k_d^{1/3}$. The values of $m_s/m_{so}$ at $r_{ef}$ proportional to $r^2$ are lower than those at $r_{ef}$ proportional to $r$. These values are given in the table for several different values of $k_o$ and $k_m$ (at $k_d = 1$). They serve as numerical examples of the fact that *larger bodies grow faster at $r_{ef}$ proportional to $r^2$*.

## MERGER OF TWO COLLIDING PREPLANETESIMALS

### *Model of the Collision of Preplanetesimals*

In order to illustrate the possibility of merger of two rarefied preplanetesimals and determine whether this possibility does not contradict our model of formation of binary objects, we consider the following simple model below: Each preplanetesimal is a sphere with diameter $D_s$ and mass $M$, which is formed by $N$ identical boulder- or particle-sized objects with diameter $d_p$. If the diameter of a solid planetesimal formed after the compression of a preplanetesimal is denoted as $D$ and the planetesimal and its constituent objects have equal densities, then $N = (D/d_p)^3$.

In the model considered below, boulders forming the second preplanetesimal move within the first homogeneous preplanetesimal along a straight line. This model provides the minimum estimate of the distance travelled by a boulder within the first preplanetesimal. In the two-body problem, a boulder actually moves prior to the collision along a section of a second-order curve. This section is longer than a section of a straight line (due to the smallness of relative velocities of encountering preplanetesimals the trajectory is even longer if the influence of the Sun is taken into account). In fact, when this boulder collides with a certain boulder of the first preplanetesimal, its trajectory changes (these boulders may also merge or break up). If a boulder in the considered model travels at a minimum distance of $0.5k_r D_s$ from the center of the first preplanetesimal along a straight line, then the length of its track within the first sphere is $L = D_s k_s = D_s(1 - k_r^2)^{1/2}$. At $0 \leq k_r < 0.9165$, the value of $L$ differs from that at $k_r = 0$ by a factor of no more than 2.5 (i.e., $L > 0.4 D_s$). This means that if an approaching boulder of the second preplanetesimal collides with several boulders of the first preplanetesimal at $k_r = 0$, then a collision of the centers of masses of two identical preplanetesimals implies that most boulders of the second preplanetesimal also collide with several boulders of the first preplanetesimal.

**Table.** Range of values of ratio $m_s/m_{so}$ between the end and the initial masses of a planetesimal satellite (the smaller value corresponds to effective radius $r_{ef}$ proportional to $r^2$, and the larger one corresponds to $r_{ef}$ proportional to radius $r$ of the satellite) at ratio $k_m = 2$ and 10 between the end and the initial masses of the planetesimal and ratio $k_o = 0.5$, 0.1, and 0.01 between the initial masses of the satellite and the planetesimal

| $k_o$ \ $k_m$ | 2 | 10 |
|---|---|---|
| 0.5 | 1.23 – 1.49 | 1.43 – 2.78 |
| 0.1 | 1.02 – 1.13 | 1.03 – 1.32 |
| 0.01 | 1.001 – 1.026 | 1.0015 – 1.057 |

If two identical spherical preplanetesimals with diameter $D_s = 2k_H r_H$ move in one plane in circular heliocentric orbits separated by a distance of $\Theta D_s$, then the volume of the part of the second sphere with boulders penetrating the first sphere is $V_h = 3^{-1}\pi h^2(1.5D_s - h)$, where $h = D_s(1 - \Theta)$. Thus, if $\Theta$ is not close to zero, not all boulders of collided preplanetesimals are included in the formed preplanetesimal, and the masses and angular momenta of the formed preplanetesimals are lower than the corresponding values in the case of a complete merger of preplanetesimals. At $\Theta = 0.5$, we have $V_h/V = 0.5$, where $V$ is the volume of a sphere with diameter $D_s$. This estimate shows that more than half the boulders of the second preplanetesimal penetrate inside the first preplanetesimal at $\Theta < 0.5$ (i.e., in 50% of preplanetesimal collisions, if we assume that colliding homogeneous preplanetesimals are distributed uniformly over $\Theta$), since $V_h/V > 0.5$ in this case.

Below in this section we assume that boulders forming both homogeneous preplanetesimals have equal diameters $d_p$, and the volume swept by a single boulder inside the other preplanetesimal is $\pi \cdot d^2 D_s k_s$, where $k_s = (1 - k_r^2)^{1/2}$. The ratio of this volume to volume of a preplanetesimal divided by $N$ yields the minimum number of collisions for a single boulder: $N_{col} = N(\pi \cdot d^2 D_s k_s)/(\pi \cdot 6^{-1} D_s^3) = 6Nd^2 k_s D_s^{-2} = 6k_s D^3/(D_s^2 d_p)$. Let us take the preplanetesimal diameter equal to $D_s = 2k_H \cdot a(M/3M_S)^{1/3}$, the density of boulders and the planetesimal formed via compression of the preplanetesimal equal to $\rho = k_\rho$ g cm$^{-3}$, and $a = k_a$ AU. Then

$$N_{col} \approx 1.5 \cdot 3^{2/3} \cdot k_s \cdot D^3 (M_S/M)^{2/3}/(a^2 d_p) = 1.5 \cdot 3^{2/3} \cdot (6/\pi)^{2/3} k_s \cdot D(M_S)^{2/3}/(a^2 d_p \cdot \rho^{2/3}) \approx$$
$$\approx 3^{-1} \cdot 10^{-3} \cdot k_a^{-2} k_\rho^{-2/3} k_s^{1} k_H^{-2} D \cdot d_p^{-1}. \tag{9}$$

Since even the "impactless" (without boulder collisions) trajectory of a boulder inside the first preplanetesimal is longer than the section of a straight line considered above, $N_{col}$ is the lower estimate of the real number of collisions $N_{col-r}$. Ratio $N_{col-r}/N_{col}$ increases with $N_{col}$, since the overall length of the path of a boulder inside a sphere increases on average with the number of its collisions with boulders of the first preplanetesimal. The minimum free path length of a boulder inside a preplanetesimal is $D_s k_s/N_{col}$. At $N_{col} \geq 1$, a boulder collides with other boulders at least once. After the first collision, boulders may merge, break up, or shift to a different trajectory (models of collisions of boulders and particles were discussed by Wahlberg Jansson et al. (2017)). The higher $N_{col}$ is (i.e., the lower the ratio of the free path length to $D_s k_s$ is), the higher is the probability that a boulder or its fragments remain inside a preplanetesimal.

It follows from (9) that $N_{col} = 1$ at, e.g., $D = 1000$ km, $d_p = 0.13$ m, $k_\rho = k_s = k_H = 1$, and $a = 40$ AU. In relation (9), $N$ is proportional to $Dk_H^{-2}$. This proportionality suggests that $N_{col}$ and the probability of merger of colliding preplanetesimals are higher at lower values of ratio $k_H$ of the radius of a preplanetesimal to its Hill radius. $N_{col}$ decreases with $D$ (i.e., more massive preplanetesimals are more likely to merge in a collision). At $D = 40$ km and $k_H = 0.2$, the value of $N_{col}$ is the same as at $D = 1000$ km and $k_H = 1$. $N_{col} \approx 10^3$ at $a = 1$ AU, $D = 1000$ km, $d_p = 0.1$ m, $k_s = k_H = 1$, and $k_\rho = 5.52$.

The lower diameter $d_p$ is (if the preplanetesimal mass remains unchanged), the higher is the probability of capture of boulders or dust particles in preplanetesimal collisions. The capture probability is higher for those boulders (particles) that travel closer to the center of the other preplanetesimal (and thus have a generally longer travel path), especially if the preplanetesimal density decreases with distance from its center. The trajectory of relative motion of centers of mass of encountering preplanetesimals may differ considerably from a parabola or an ellipse (motion trajectories in the two-body problem) and may be fairly complex with several revolutions around the common center of mass (Ipatov, 1987). Therefore, the path length of a boulder inside the other preplanetesimal (even before the first collision occurs) may sometimes exceed $D_s$ considerably, and $k_s$ may be several times higher than unity.

*Relative Velocities of Colliding Preplanetesimals*

The relative velocities of colliding boulders during encounters of preplanetesimals may be very low. Let us discuss the case when the ratios of radii $r_1$ and $r_2$ of colliding preplanetesimals to their Hill radii are equal to $k_H$, and these preplanetesimals moved in circular heliocentric orbits prior to the collision. Using formula (3), we then determine that the ratio of tangential component $v_\tau$ of the preplanetesimal collision velocity to parabolic velocity $v_{par}$ (proportional to $r_1^{-1/2}$) on the surface of the larger preplanetesimal with radius $r_1$ is $v_\tau/v_{par}=k_\Theta \cdot k_H^{3/2} \cdot 6^{-1/2}$ at $r_1 \gg r_2$ and $v_\tau/v_{par}=2k_\Theta \cdot k_H^{3/2} \cdot 6^{-1/2}$ at $r_1=r_2$. Since $k_\Theta \cdot 6^{-1/2} \leq 0.4$ at $k_\Theta \approx (1-1.5\Theta^2) \leq 1$, the tangential preplanetesimal collision velocity is lower than the parabolic velocity $v_{par}$. The maximum values of collision velocity $v_{col}$ (see formula (2)) and $v_\tau$ are equal. These velocity estimates suggest that colliding preplanetesimals may merge.

The above estimates of the ratio $v_\tau/v_{par}$ did not take into account the mutual gravitational influence of preplanetesimals, which becomes significant at smaller $k_H$. If the heliocentric orbits of both preplanetesimals, radii $r$ of which were much shorter than their Hill radii $k_H$, were almost circular before they came within their Hill sphere, then it will be shown at the end of this paragraph that preplanetesimal collision velocity is produced largely by an increase in the relative velocity of preplanetesimals during their motion inside the Hill sphere (and not by the velocity of approach to within $r_H$), and $v_{col}$ is close to $v_{par}$. Therefore, the collision of rarefied preplanetesimals may result in their merger (with probable subsequent formation of a satellite system) at any $r_1 < r_H$ (not only at $r_1 \approx r_H$). In contrast to the above estimates, the motion of preplanetesimals inside the sphere with its radius being larger than the sum of radii of two encountering preplanetesimals is considered in the present paragraph. If the bodies (rarefied preplanetesimals or solid planetesimals) are assumed to have moved in circular heliocentric orbits before they came within radius $r_{en}=\beta \cdot r_H$ ($\beta \leq 1$) of each other, then one may demonstrate that the ratio of their relative velocity at the moment of this close approach to parabolic $v_{par}$ on the surface of the larger body is approximately equal to $6^{-1/2} \cdot \beta \cdot l_\theta \cdot \alpha_r^{1/2}$, where $\alpha_r$ is the ratio of the sum of radii of these bodies to radius $r_{en}$ of the considered sphere. It follows from (2) that the ratio of relative velocity $v_{sH}$ of bodies coming within $r_{en}$ of each other to the velocity of a body in a circular heliocentric orbit with radius $a$ is approximately equal to $(1 - 0.75\Theta^2)^{1/2} r_{en}/a$, where the difference between semimajor axes of heliocentric orbits of the bodies approaching each other is $\Theta \cdot r_{en}$. If the Hill sphere is considered and $\Theta = 1$, then $v_{sH}/v_{par} \approx 0.2\alpha_r^{1/2}$ and, consequently, $v_{sH} < v_{par}$. For smaller spheres, the values of $v_{sH}$ are lower. Therefore, parabolic velocity $v_{par}$ in the considered model is higher than velocity $v_{sH}$ of encounter to $r_{en}$ at any $\alpha_r$ ($\alpha_r \leq 1$). Analyzing the two-body one finds that $v_{col}^2 \approx v_{sH}^2 + v_{par}^2$ at $\alpha_r \ll 1$. The results of calculations made by Ipatov (1981a; 1981b) have demonstrated that $v_{col}$ is close to $v_{par}$ if $e_m m_r \leq 5$, where $e_m$ is the maximum eccentricity of the heliocentric orbits of two colliding bodies, and $m_r$ is the mass of the larger body (expressed in solar masses). Multiple (hundreds of thousands in each run) motions of bodies inside the considered sphere were modeled in these calculations for various masses and elements of heliocentric orbits of these bodies (before their entry into the sphere). The probabilities and parameters of collisions and the $\Delta K$ values were calculated.

*Discussion of Models of Merger of Colliding Preplanetesimals*

The above estimates suggest that *a boulder or a particle belonging to the second preplanetesimal, which moves inside the first one, may be captured by the first preplanetesimal (at least if the diameter of this boulder or particle is sufficiently small).* Although not all collisions of preplanetesimals end in their merger (this is especially true for grazing collisions), it is fair to assume that a considerable number of collisions resulted in the formation of fairly large preplanetesimals with an angular momentum sufficient to form satellite systems. It was noted in (Ipatov, 2017) that the model of formation of binary objects in preplanetesimal colli- sions explains the observed distributions of orbital elements of the second components of binary TNOs (including the negative angular momenta of some of these objects).

It is known that the presence of gas may enhance the probability of capture of small objects or particles, which compose preplanetesimals, in preplanetesimal collisions (and thus make the formed preplanetesimal more massive), since gas may reduce the eccentricities of heliocentric orbits of preplanetesimals and, consequently, reduce the relative velocities of colliding pre planetesimals and raise the probabilities of their mergers. Nesvorny et al. (2010) have concluded that gas drag does not exert a strong influence on the formation of a binary object via condensation compression. Therefore, it is fair to assume that the inclusion of the influence of gas should not alter the conclusions regarding the significance of preplanetesimal collisions and their contribution to the formation of binary TNOs.

The probable mergers of rarefied preplanetesimals do not contradict the results obtained in several earlier studies. For example, collided preplanetesimals did merge in the calculations performed by Johansen et al. (2011). In the numerical model of Johansen et al. (2007), the average free path length of a boulder inside a condensation (preplanetesimal) is shorter than the condensation diameter. Lyra et al. (2008) have noted that the velocity dispersion of rarefied preplanetesimals did not exceed 1 m s$^{-1}$ in most calculations; therefore, disruptive collisions of boulders did not occur. In certain models of evolution of the rings of Saturn (R. Perrine, private communication, 2009), colliding objects (ring elements) form a new object if their collision velocity is lower than the mutual escape velocity by a certain factor. Several arguments in favor of merger of colliding preplanetesimals were presented in (Ipatov, 2010).

The mass of a condensation formed in a collision of two preplanetesimals may be lower than the sum of masses of these preplanetesimals. It follows from Fig. 2 in (Nesvorny et al., 2010) that the mass of the formed solid binary object was more than five times lower than the mass of its parental condensation. Therefore, *the mass of a binary object may sometimes be several times lower than the sum of masses of colliding preplanetesimals*. However, under certain conditions, the mass of the formed binary object may be close to the sum of masses of colliding preplanetesimals. For example, Galimov (2011) and Galimov and Krivtsov (2012) have proposed a model in which the Earth–Moon system is produced via compression of a condensation with its angular momentum and mass being equal to the angular momentum and the mass of this system. The angular momentum of a single TNO may be significantly (e.g., two orders of magnitude) lower than the angular momentum of a satellite system of the same mass. However, the results of calculations presented in (Nesvorny et al., 2010) suggest that it is also possible for the angular momenta of parental conden- sations of a satellite system and a single object to differ only slightly (e.g., by a factor of less than two).

## FRACTION OF RAREFIED PREPLANETESIMALS THAT MAY COLLIDE WITH OTHER PREPLANETESIMALS

*Model for Estimating the Number of Collisions of Preplanetesimals*

The process of the preplanetesimal disk evolution was a complex one. For example, different preplanetesimals could form and contract at different times and have different sizes, their radii could decrease with time, the diameters of certain preplanetesimals could be smaller than the disk thickness (i.e., it is possible that the preplanetesimal disk was not ideally flat), etc. Still, a simple model of the rate of preplanetesimal collisions in the disk is considered below, and rough estimates of the fraction of preplanetesimals colliding with other preplanetesimals of a close size are given for this model. Such estimates may serve as the basis for the analysis of other disk models. The estimated times of compression of preplanetesimals (condensations) given in the studies mentioned in the Introduction did differ strongly and varied from 25 years to millions of years. We will determine below which preplanetesimal compression times are a better fit to the model of formation of trans-Neptunian satellite systems in preplanetesimal collisions.

Since the heliocentric orbits of preplanetesimals were almost circular, the colliding preplanetesimals should have formed at almost equal distances from the Sun. Therefore, to estimate the number of preplanetesimal collisions, one may consider a very narrow disk with a width of several Hill radii of preplanetesimals. The range of preplanetesimal formation times at a fixed distance from the Sun may be narrower than that for a wider disk.

Let us consider the number of collisions of preplanetesimals with masses $m_o = 6 \times 10^{17}$ kg $\approx 10^{-7} M_E$ (e.g., with masses of solid bodies with diameter $d_s = 100$ km and density $\rho \approx 1.15$ g cm$^{-3}$; $M_E$ is the mass of the Earth) moving in one and the same plane. The Hill radius of such a preplanetesimal is

$r_{Ho} \approx 4.6 \times 10^{-5}a$. In the case of circular heliocentric orbits with the difference between their semimajor axes being equal to Hill radius $r_{Ho}$, the ratio of periods of motion of two preplanetesimals around the Sun is approximately $1+1.5r_{Ha} \approx 1 + 7 \times 10^{-5}$, where $r_{Ha} = r_{Ho}/a$. The angle with its vertex at the Sun and its sides pointing at two preplanetesimals then varies by $2\pi 1.5 r_{Ha} n_r \approx 0.044$ radians in $n_r = 100$ revolutions of these preplanetesimals around the Sun. Let us assume that a preplanetesimal may collide with another preplanetesimal if the semimajor axes of their orbits differ by no more than $2r_{Ho}$. We consider a disk with ratio $a_{rat} = 1.67$ of the distances between its edges and the Sun (e.g., a disk at 30–50 AU from the Sun), where the shorter distance is $a_{min}$. If its surface density is constant, the number of planetesimals found within the range of solar distances from $a-2r_H$ to $a+2r_H$ is $N_{mH}=8N \cdot r_{Ha}(a/a_{min})^2/(a_{rat}^2-1)$, where $N$ is the number of preplanetesimals in the disk, and $r_{Ha}$ and $a_{ra}$ are dimensionless quantities. $N=10^7$ at $m_o=10^{-7}M_E$ and total disk mass $M_\Sigma = M_E$ (the mass of the Earth). If this is the case, $N_{mH}$ varies from $2.5 \times 10^3$ to $6.9 \times 10^3$ for the inner and the outer edges of the disk, respectively. Average number $N_c$ of collisions of the considered preplanetesimal with other preplanetesimals in $n_r$ revolutions around the Sun may be estimated as $1.5r_{Ha}n_r N_{mH}$ (if we assume that collisions may occur when the semimajor axes of encountering preplanetesimals differ by no more than $2r_{Ha}$). It was taken into account here that the relative angular velocity of two preplanetesimals with the difference between their semimajor axes being equal to $r_{Ho}$ is $3\pi r_{Ha}$, while the periods of their revolution around the Sun differ by a factor of $1 +1.5r_{Ha}$. The probability of a collision between such preplanetesimals in a single revolution around the Sun is $3\pi r_{Ha}/2\pi = 1.5 r_{Ha}$. At $r_{Ha} \approx 4.6 \times 10^{-5}$ and under the assumption that preplanetesimals may collide if their semimajor axes differ by no more than $2r_{Ho}$, average number $N_{c1} = N_c/n_r$ of preplanetesimal collisions in a single revolution around the Sun is 0.2 at $N_{mH} \approx 3 \times 10^3$ and 0.4 at $N_{mH} \approx 6 \times 10^3$. Note that $N_c/n_r$ is proportional to $r_{Ha}N$ (one factor $r_{Ha}$ stems from the size of the region where condensations may collide, and the other factor $r_{Ha}$ is established by the relative angular velocity of condensations) and is also proportional to $M_\Sigma \cdot m_o^{-1/3}$ and $M_\Sigma \cdot d_s^{-1}$. At $M_\Sigma=10 M_E$ and $d_s= 1000$ km (i.e., at $N = 10^5$ and $\rho \approx 1.15$ g cm$^{-3}$), $N_c$ is the same as for $M_\Sigma = M_E$ and $d_s = 100$ km (i.e., at $N = 10^7$ and $\rho \approx 1.15$ g cm$^{-3}$). The times to collisions between certain preplanetesimals could be shorter than the average ones, since the initial distances between them could also be short.

*Discussion of the Rate of Preplanetesimal Collisions*

The above rough model may be used to estimate the number of preplanetesimal collisions in a disk with different parameters. The actual rate of preplanetesimal collisions is much lower than these estimates. For example, if the semimajor axes of orbits of colliding preplanetesimals differ by no more than $r_{Ho}$ (instead of $2r_{Ho}$), then collision rate $N_c/n_r$ is four times lower than the above estimates obtained for $2r_{Ho}$ (the number of collision candidates is two times lower in this case, and the characteristic time to a collision is longer). In our view, the ratio of the number of such preplanetesimals that were formed in collisions of preplanetesimals and acquired angular momenta sufficient for the formation of trans-Neptunian satellite systems to the total number of TNO-producing preplanetesimals is close to 0.45 (the initial fraction of classical TNOs with satellites among the discovered classical TNOs (Petit and Mousis, 2004)).

Preplanetesimals contracted over time, and the radii of many collided preplanetesimals could be sig- nificantly less than their Hill radii. The preplanetesimal disk could be nonplanar. If radius $r_p$ of a preplanetesimal is less than the Hill radius and equals $k_H r_H$, then the number of encounters between preplanetesimals within the Hill sphere does not change, but the number of collisions inside the Hill sphere decreases with a reduction in $k_H$. At low relative velocities of preplanetesimals (i.e., in the case of almost circular orbits of preplanetesimals), the effective preplanetesimal radius is proportional to $r_p^2$ due to gravitational focusing. The collision probability in the planar model is then proportional to $k_H^2$, and the effective gravitational cross section and the collision probability in the three-dimensional model are proportional to $r_p^4$ and $k_H^4$. In the case of high relative velocities, the effective radius is close to $k_H r_H$, and the effective gravitational cross section in the three-dimensional model is close to the preplanetesimal cross section. If the preplanetesimal diameter is smaller than the preplanetesimal disk thickness and the planar model is not applicable, the dependence of the collision probability on $r_p$ is stronger than that in the planar model.

Certain preplanetesimal collisions were grazing and did not result in the merger of preplanetesimals. Not all mergers led to the formation of a preplanetesimal with an angular momentum sufficient to form a satellite system. Preplanetesimals could form at different times, and the number of preplanetesimal present at a given time moment could be lower than the one in the above model with the same disk mass. These factors reduce the preplanetesimal collision rate relative to the model discussed in previous subsection. On the other hand, the preplanetesimal disk could be more massive than in the model described above, which translates into an increased collision rate. The time of preplanetesimal compression in certain models men- tioned in the Introduction exceeds 1000 revolutions around the Sun. In the case of eccentric preplanetesimal orbits, the number of candidates for a collision with a given preplanetesimal is higher than that in the case of circular orbits, but high collision velocities translate into a probable reduction in the mass of a preplanetesimal formed after a collision and in the probability of a merger of collided preplanetesimals. Thus, the condition necessary for the formation of the observed number of binary TNOs may be fulfilled in considering the collisions of preplanetesimals that were formed and existed at different times and had typical radii less than their Hill radii. As was already noted, this condition for the considered model consists in that the fraction of preplanetesimals colliding with other preplanetesimals and forming new preplanetesimals with angular momenta sufficient to form trans-Neptunian satellite systems should be equal to the initial fraction of actual trans-Neptunian satellite systems among all TNOs. The above reasoning suggests that the preplanetesimal compression times determined in different studies are not incompatible with the model of formation of binary TNOs via compression of parental preplanetesimals with the major fractions of their angular momenta acquired in preplanetesimal collisions.

## FORMATION OF TRANS-NEPTUNIAN OBJECTS AND THEIR SATELLITES AT DIFFERENT DISTANCES FROM THE SUN

The formation of satellite systems of TNOs at different distances from the Sun is discussed in the present section based on the results presented in the previous sections. The studies focused on the issue of formation of satellites of minor bodies were reviewed in the Introduction. Except for (Ipatov, 2010; Nesvorny et al., 2010), all these studies were concerned with the formation of satellites at the stage when minor bodies were solid (and not at the earlier stage of rarefied preplanetesimals).

### *Estimates of the Dependence of the Fraction of Satellite Systems on the Distance from the Sun to Their Formation Site*

We assume that the fraction of preplanetesimals colliding (with a necessary angular momentum) and merging with other preplanetesimals of similar sizes in the process of their compression could be close to the initial fraction of minor bodies with satellites; i.e., it could be equal to 0.45 for objects with diameter $d_s > 100$ km formed in the trans-Neptunian belt (Petit and Mousis, 2004) and could be lower in the asteroid belt. The fact that the number of satellite systems in the main asteroid belt is lower than the corresponding number in the trans-Neptunian belt suggests that most rarefied protoasteroids could turn into solid asteroids before colliding with other protoasteroids of a similar size. Certain currently observed asteroids (especially with $d_s < 10$ km) are believed to be fragments of larger solid bodies.

In the model discussed in the first section, angular velocity ω of a preplanetesimal formed in a collision of preplanetesimals is proportional to $a^{-3/2}$. If the radii of preplanetesimals are proportional to their Hill radii (i.e., proportional to semimajor axis *a*), then angular velocity $\omega_s$ of a solid planetesimal formed via compression of a preplanetesimal is $\omega \cdot k_r^{-2}$ and is proportional to $a^{1/2}$, where $k_r$ is the ratio of the radius of a solid planetesimal to the radius of its parental rarefied preplanetesimal. In the case of fast rotation, not all material of the contracting preplanetesimal is actually transferred into the formed solid planetesimal. One of the probable reasons for a major fraction of binary objects to form at long distances $a$ from the Sun is as follows: preplanetesimals were larger at long $a$, and a larger fraction of material of a rotating contracting preplanetesimal could enter a cloud surrounding the dense core at the preplanetesimal center (or two compression centers were more likely to exist).

Let us consider the formation of a new preplanetesimal in a collision of preplanetesimals with masses $m_1$ and $m_2$ ($m_1 \geq m_2$) and radii proportional to their Hill radii $r_H$. Since $r_H$ is proportional to $am^{1/3}$, $J_S = 0.4\chi(m_1 + m_2)r^2$ ($r$ is the radius of the formed preplanetesimal), and $T_S = 2\pi J_S/K_S$, one may determine using formula (1) that the period of axial rotation of the object formed as a result of compression of the new preplanetesimal to rdius $r_c$ is proportional to $r_c^2 \cdot a^{-1/2} \cdot (m_1+m_2)^2 \cdot m_1^{-1} \cdot m_2^{-1} \cdot (m_1^{1/3}+m_2^{1/3})^{-2}$. Let us analyze this proportionality at a fixed $r_c$. If $a$ increases, the critical period of axial rotation of the contracting preplanetesimal (when the velocity of a particle on the surface of this preplanetesimal exceeds the escape velocity) may be attained at a higher ratio $m_1/m_2$ of masses of colliding preplanetesimals and a lower value of $m_2$. At $m_1=m_2$, $T_S$ is proportional to $r_c^2 \cdot a^{-1/2} m_1^{-2/3}$. It also follows from this proportionality that the critical period value may be attained at lower masses of bodies if $a$ increased. These are the arguments in favor of the hypothesis that a preplanetesimal disk more distant from the Sun has, under otherwise equal conditions, a higher fraction of preplanetesimal collisions (for a larger probable difference in the masses of collided preplanetesimals and for lower preplanetesimal masses) that may lead to the formation of binary objects.

The above reasoning suggests that if the considered binary objects have formed from rarefied preplanetesimals produced in preplanetesimal collisions (and not, for example, in collisions of solid bodies), then the fraction of binary objects (or minor bodies with several satellites) increases, under otherwise equal conditions, with distance from the Sun to the region of formation of parental preplanetesimals.

*Estimates of the Dependence of the Fraction of Satellite Systems on the Preplanetesimal Disk Mass*

Total mass $M_\Sigma$ of preplanetesimals in their feeding zone also affects the fraction of binary objects. In a planar disk consisting of identical preplanetesimals, which was discussed in the previous section, the number of preplanetesimals is proportional to $M_\Sigma/d^3$, and average number $N_c$ of collisions between a preplanetesimal with diameter $d$ and other preplanetesimals is proportional to $M_\Sigma/d$ in certain cases. This proportionality is established if probability $p$ of a single collision is proportional to $d^2$, which is true, for example, for the planar model with effective radius $r_{ef}$ of a preplanetesimal being shorter than the Hill radius and proportional to $d^2$. The same $p$ proportionality is valid for Hill spheres if the disk thickness is not lower than the Hill radius and if preplanetesimals move in unperturbed circular orbits outside Hill spheres. In the case of solid planetesimals moving in strongly inclined and eccentric orbits, probability $p$ of a single collision is proportional to $r_{ef}$, and $r_{ef}$ is proportional to $d$; therefore, $N_c$ is also proportional to $M_\Sigma/d$. At small inclinations and eccentricities of orbits of preplanetesimals (smaller than their Hill spheres), $p$ may be proportional to $r_{ef}^2$, and $r_{ef}$ may be proportional to $d^2$; as a result, $N_c$ is proportional to $M_\Sigma \cdot d$.

The values of $M_\Sigma/d$ and $M_\Sigma d$ in the asteroid belt were probably several times lower than those in the trans-Neptunian belt (in the case of a disk composed of different preplanetesimals, d may be regarded as a typical diameter). Lower $M_\Sigma/d$ and $M_\Sigma d$ values correspond to lower numbers of collisions between preplanetesimals and lower fractions of binary objects formed from preplanetesimals in the asteroid belt (relative to the trans-Neptunian belt). In the model from the previous section, preplanetesimals move in one plane. If their diameters $d$ are smaller than the disk thickness (this could be true for condensations producing small asteroids), preplanetesimal collisions occur less frequently than in the planar model, and the fraction of formed satellite systems is thus lower. Our model covers only the formation of satellite systems via condensation compression and provides no explanation of the origin of binary asteroids that are fragments of large asteroids and TNOs. The fact that the fraction of TNOs with satellite systems in more eccentric orbits (these objects are believed to have arrived from the Uranus and Neptune feeding zone) is lower (relative to classical TNOs) may be attributed to lower probabilities of a collision and merger of two collided preplanetesimals in this region. These probabilities, in their turn, are lower than in the trans-Neptunian belt due possibly to larger eccentricities e and inclinations i of preplanetesimal orbits in the Uranus and Neptune feeding zone. Larger $e$ and $i$ values may result from the abundance of material in the feeding zone of the giant planets.

*Discussion of Formation of Trans-Neptunian Objects and Their Satellites*

The above hypothesis regarding the increased fraction of binary objects formed at a greater distance from the Sun was formulated under the assumption that the time (expressed in revolutions around the Sun) from the moment of formation of planetesimals to their collision for TNOs is not shorter than (e.g., equal to) the corresponding time for asteroids. In most studies mentioned in the Introduction, the times of formation of preplanetesimals (in revolutions around the Sun) did not depend on the distance to the Sun.

The formation of classical TNOs from rarefied preplanetesimals could be accomplished with a present total mass of preplanetesimals in the trans-Neptunian region (even with the current total mass of TNOs). The models of formation of TNOs by accumulation of solid planetesimals (see, for example, Stern, 1995; Kenyon and Luu, 1998; 1999) require a massive initial belt (with its total mass exceeding ten Earth masses) and low (~0.001) eccentricities in the process of accumulation. However, the results of calculations (e.g., Ipatov, 2007) demonstrate that the gravitational interaction between planetesimals at this stage could raise their eccentricities to such values that exceed considerably the mentioned value of ~0.001. This increase of eccentricities provides evidence against the model of prevalent growth of TNOs via accumulation of small planetesimals and, consequently, supports the model of formation of large TNOs from rarefied preplanetesimals.

CONCLUSIONS

Trans-Neptunian objects (including the ones with satellites) could form as a result of compression of rarefied preplanetesimals. The angular velocities used by Nesvorny et al. (2010) as the initial data for modeling the compression of rarefied preplanetesimals resulting in the formation of binary TNOs could be acquired in collisions of rarefied preplanetesimals with their sizes being close to their initial sizes.

The contribution of a collision of two identical rarefied preplanetesimals to the angular momentum of the produced rarefied preplanetesimal may be larger (up to 12 times larger) than the contribution of the initial rotation of homogeneous rarefied preplanetesimals if the radii of colliding rarefied preplanetesimals differ from their initial radii by a factor of no more than three. This difference in radii may be larger in the case of inhomogeneous preplanetesimals.

Some rarefied preplanetesimals formed because of a collision of preplanetesimals in the region of formation of solid small bodies acquired such angular momenta that are sufficient to form satellite systems of small bodies. It is likely that the ratio of the number of rarefied preplanetesimals with such angular momenta to the total number of rarefied preplanetesimals producing classical TNOs with diameter $d_s > 100$ km was equal to 0.45 (the initial fraction of satellite systems among all classical TNOs).

The major part of the angular momentum of the majority of rarefied condensations producing binary TNOs was not acquired via accumulation of small objects; if this were not the case, binary TNOs would have only positive angular momenta.

ACKNOWLEDGMENTS

The author would like to thank the referees for helpful remarks. This study was supported in part by the Russian Foundation for Basic Research (projects no. 14-02-00319 and no 17-02-00507 A) and Program no. 7 of the Presidium of the Russian Academy of Sciences.

REFERENCES

Astakhov, S.A., Lee, E.A., and Farrelly, D., Formation of Kuiper-belt binaries through multiple chaotic scattering encounters with low-mass intruders, *Mon. Not. R. Astron. Soc.*, 2005, vol. 360, pp. 401–415.

Carrera, D., Johansen, A., and Davies, M.B., How to form planetesimals from mm-sized chondrules and chondrule aggregates, *Astron. Astrophys.*, 2015, vol. 579, art. ID A43, 20 p. http://arXiv.org/abs/1501.05314.

Chambers, J.E., Planetesimal formation by turbulent concentration, *Icarus*, 2010, vol. 208, pp. 505–517.

Chiang, E. and Youdin, A., Forming planetesimals in solar and extrasolar nebulae, *Ann. Rev. Earth Planet. Sci.*, 2010, vol. 38, pp. 493–522.

Ćuk, M., Formation and destruction of small binary asteroids, *Astrophys. J.*, 2007, vol. 659, pp. L57–L60.

Cuzzi, J.N. and Hogan, R.C., Primary accretion by turbulent concentration: The rate of planetesimal formation and the role of vortex tubes, *The 43rd Lunar and Planetary Science Conf., Abstracts of Papers,* Houston, TX: Lunar Planet. Inst., 2012, no. 2536.

Cuzzi, J.N., Hogan, R.C., and Shariff, K., Toward planetesimals: dense chondrule clumps in the protoplanetary nebula, *Astrophys. J.*, 2008, vol. 687, pp. 1432–1447.

Cuzzi, J.N., Hogan, R.C., and Bottke, W.F., Towards initial mass functions for asteroids and Kuiper belt objects, *Icarus*, 2010, vol. 208, pp. 518–538.

Davidson, B.J.R., Sierks, H., Güttler, C., Marzari, F., Pajola, M., Rickman, H., A'Hearn, M.F., Auger, A.-T., El-Maarry, M.R., Fornasier, S., et al., The primordial nucleus of comet 67P/Churyumov–Gerasimenko, *Astron. Astrophys.*, 2016, vol. 592, art. ID A63, 30 p.

Eneev, T.M., Evolution of the outer solar system-possible structure beyond Neptune, *Sov. Astron. Lett.* 1980, vol. 6, pp. 163-166.

Eneev, T.M. and Kozlov, N.N., A model of the accumulation process in the formation of planetary systems. II— Rotation of the planets and the relation of the model to the theory of gravitational instability, *Sol. Syst. Res.*, 1982, vol. 15, pp. 97–104.

Eneev, T.M. and Kozlov, N.N., *The Dynamics of Planet Formation. Theory and Computer Simulation*, Saar- brucken: LAP Lambert Academic, 2016, 132 p.

Fulle, M., Corte, V.D., Rotundi, A., Rietmeijer, F.J.M., Green, S.F., Weissman, P., Accolla, M., Colangeli, L., Ferrari, M., Ivanovski, S., et al., Comet 67P/Churyumov–Gerasimenko preserved the pebbles that formed planetesimals, *Mon. Not. R. Astron. Soc.*, 2016, vol. 462, pp. S132–S137.

Funato, Y., Makino, J., Hut, P., Kokubo, E., and Kinoshita, D., The formation of Kuiper-belt binaries through exchange reactions, *Nature*, 2004, vol. 427, pp. 518–520.

Galimov, E.M., Formation of the Moon and the Earth from a common supraplanetary gas-dust cloud (lecture presented at the XIX All-Russia symposium on isotope geochemistry on November 16, 2010), *Geochem. Int.*, 2011, vol. 49, no. 6, pp. 537–554.

Galimov, E.M. and Krivtsov, A.M., *Origin of the Moon. New Concept*, Berlin: De Gruyter. Verlag, 2012, 168 p.

Gibbons, P.G., Rice, W.K.M., and Mamatsashvili, G.R., Planetesimal formation in self-gravitating disks, *Mon. Not. R. Astron. Soc.*, 2012, vol. 426, pp. 1444–1454.

Giuli, R.T., On the rotation of the Earth produced by gravitational accretion of particles, *Icarus*, 1968, vol. 8, pp. 301–323.

Goldreich, P. and Ward, W.R., The formation of planetesimals, *Astrophys. J.*, 1973, vol. 183, pp. 1051–1061.

Goldreich, P., Lithwick, Y., and Sari, R., Formation of Kuiper-belt binaries by dynamical friction and three-body encounters, *Nature*, 2002, vol. 420, pp. 643–646.

Gomes, R.S., The origin of the Kuiper Belt high-inclination population, *Icarus*, 2003, vol. 161, pp. 404–418.

Gomes, R.S., On the origin of the Kuiper belt, *Celest. Mech.*, 2009, vol. 104, pp. 39–51.

Gorkavyi, N.N., Regular mechanism of the formation of asteroid satellites, *The 10th Asteroids, Comets, Meteors Meeting, Abstracts of Papers*, Baltimore: Johns Hopkins Univ., 2008, no. 8357.

Gundlach, B., Blum, J., Keller, H.U., and Skorov, Y.V., What drives the dust activity of comet 67P/Churyumov–Gerasimenko? *Astron. Astrophys.*, 2015, vol. 583, art. ID A12, 8 p. https://arxiv.org/abs/1506.08545.

Ipatov, S.I., Numerical studies of the angular momenta of accumulating bodies, *Preprint of Inst. of Applied Mathematics, Acad. Sci. Sov. Union*, Moscow, 1981a, no. 101, 28 p.

Ipatov, S.I., The formation of axial rotations of planets, *Preprint of Inst. of Applied Mathematics, Acad. Sci. Sov. Union*, Moscow, 1981b, no. 102, 28 p.

Ipatov, S.I., Accumulation and migration of the bodies from the zones of giant planets, *Earth, Moon, Planets*, 1987, vol. 39, pp. 101–128.

Ipatov, S.I., *Migratsiya nebesnykh tel v Solnechnoi sisteme (Migration of Celestial Bodies in the Solar System)*, Moscow: URSS Editorial, 2000, 320 p. http://www.rfbr.ru/rffi/ru/books/o_29239.

Ipatov, S.I., Formation of trans-Neptunian objects, *The 32nd Lunar and Planetary Science Conf., Abstracts of Papers*, Houston, TX: Lunar Planet. Inst., 2001, no. 1165.

Ipatov, S.I., Formation and migration of trans-Neptunian objects, *Proc. 14th Annual Astrophysical Conf. "The Search for Other Worlds," October 13–14, 2003*, Holt, S.S. and Deming, D., Eds., College Park, MD: Am. Inst. Phys., 2004, vol. 713, pp. 277–280. http://arXiv.org/format/astro-ph/0401279.

Ipatov, S.I., Growth of eccentricities and inclinations of planetesimals due to their mutual gravitational influence, *The 38th Lunar and Planetary Science Conf., Abstracts of Papers*, Houston, TX: Lunar Planet. Inst., 2007, no. 1260.

Ipatov, S.I., Formation of binaries at the stage of rarefied preplanetesimals, *The 40th Lunar and Planetary Science Conf., Abstracts of Papers*, Houston, TX: Lunar Planet. Inst., 2009, no. 1021.

Ipatov, S.I., The angular momentum of two collided rarefied preplanetesimals and the formation of binaries, *Mon. Not. R. Astron. Soc.*, 2010, vol. 403, pp. 405–414. http://arxiv.org/abs/0904.3529.

Ipatov, S.I., Angular momenta of collided rarefied preplanetesimals, *Proc. Int. Astronomic Union Symp. "Formation, Detection, and Characterization of Extrasolar Habitable Planets,"* Haghighipour, N., Ed., Cambridge: Cambridge Univ. Press, 2014, vol. 8, no. 293, pp. 285–288. http://arxiv.org/abs/1412.8445.

Ipatov, S.I., The role of collisions of rarefied condensations in formation of embryos of the Earth and the Moon, *The 46th Lunar and Planetary Science Conf., Abstracts of Papers,* Houston, TX: Lunar Planet. Inst., 2015a, no. 1355.

Ipatov, S.I., The Earth-Moon system as a typical binary in the Solar System, *Int. Space Forum "SpaceKazan-IAPS-2015,"* Marov, M.Ya., Ed., Kazan: Kazan. Gos. Univ., 2015b, pp. 97–105. http://arxiv.org/abs/1607.07037.

Ipatov, S.I., Origin of orbits of secondaries in discovered trans-Neptunian binaries, *The 46th Lunar and Planetary Science Conf., Abstracts of Papers,* Woodlands, TX, 2015c, no. 1512.

Ipatov, S.I., Formation of celestial bodies with satellites at the stage of rarefied condensations, *Mater. Chetvertogo mezhdunarod. simpoziuma po issledovaniyu Solnechnoi sistemy, nauchnaya sessiya posvyashchennaya 80-letiyu akademika M.Ya. Marova "Issledovaniya Solnechnoi sistemy: Kosmicheskie vekhi"* (Proc. Forth Moscow Solar System Symp., 4M-S$^3$, Academician M. Marov 80th Anniversary Session "Solar System Study: Some Milestones"), Zakharov, A.V., Ed., Moscow: Inst. Kosm. Issled., Ross. Akad. Nauk, 2015d, vol. 7, no. 3 (56), pp. 386–399. http://www.iki.rssi.ru/books/2015marov.pdf.

Ipatov, S.I., Origin of orbits of secondaries in discovered trans-Neptunian binaries, *Sol. Syst. Res.*, 2017, vol. 51, no. 5, 409-416.

Johansen, A. and Lacerda, P., Prograde rotation of protoplanets by accretion of pebbles in a gaseous environment, *Mon. Not. R. Astron. Soc.*, 2010, vol. 404, pp. 475–485.

Johansen, A., Oishi, J.S., Mac Low, M.-M., Klahr, H., Henning, T., and Youdin, A., Rapid planetesimal formation in turbulent circumstellar disks, *Nature*, 2007, vol. 448, pp. 1022–1025.

Johansen, A., Youdin, A., and Klahr, H., Zonal flows and long-lived axisymmetric pressure bumps in magnetorotational turbulence, *Astrophys. J.*, 2009a, vol. 697, pp. 1269–1289.

Johansen, A., Youdin, A., and Mac Low, M.-M., Particle clumping and planetesimal formation depend strongly on metallicity, *Astrophys. J.*, 2009b, vol. 704, pp. L75– L79.

Johansen, A., Klahr, H., and Henning, T., High-resolution simulations of planetary formation in turbulent protoplanetary discs, *Astron. Astrophys.*, 2011, vol. 529, art. ID A62, 16 p.

Johansen, A., Youdin, A.N., and Lithwick, Y., Adding particle collisions to the formation of asteroids and Kuiper belt objects via streaming instabilities, *Astron. Astro- phys.*, 2012, vol. 537, art. ID A125, 17 p.

Johansen, A., Jacquet, E., Cuzzi, J.N., Morbidelli, A., and Gounelle, M., New paradigms for asteroid formation, in *Asteroids IV, Space Sci. Ser.*, Michel, P., DeMeo, F., and Bottke, W., Eds., Tucson, AZ: Univ. Arizona Press, 2015a, pp. 471–492. http://arxiv.org/abs/1505.02941.

Johansen, A., Mac Low, M.-M., Lacerda, P., and Bizzarro, M., Growth of asteroids, planetary embryos, and Kuiper belt objects by chondrule accretion, *Sci. Adv.*, 2015b, vol. 1, no. 3, art. ID e1500109, 11 p. http://arxiv.org/abs/1503.07347.

Kenyon, S.J. and Luu, J.X., Accretion in the early Kuiper belt. I. Coagulation and velocity evolution, *Astron. J.*, 1998, vol. 115, pp. 2136–2160.

Kenyon, S.J. and Luu, J.X., Accretion in the early Kuiper belt. II. Fragmentation, *Astron. J.*, 1999, vol. 118, pp. 1101–1119.

Kretke, K.A. and Levison, H.F. Evidence for pebbles in comets, *Icarus*, 2015, vol. 262, pp. 9–13. https://arxiv.org/abs/1509.00754.

Lambrechts, M. and Johansen, A., Rapid growth of gas-giant cores by pebble accretion, *Astron. Astrophys.*, 2012, vol. 544, art. ID A32, 13 p.

Levison, H.F. and Stern, S.A., On the size dependence of the inclination distribution of the main Kuiper belt, *Astron. J.*, 2001, vol. 121, pp. 1730–1735.

Lyra, W., Johansen, A., Klahr, H., and Piskunov, N., Embryos grown in the dead zone. Assembling the first protoplanetary cores in low mass self-gravitating circumstellar disks of gas and solids, *Astron. Astrophys.*, 2008, vol. 491, pp. L41–L44.

Lyra, W., Johansen, A., Zsom, A., Klahr, H., and Piskunov, N., Planet formation bursts at the borders of the dead zone in 2D numerical simulations of circumstellar disks, *Astron. Astrophys.*, 2009, vol. 497, pp. 869–888.

Makalkin, A.B. and Ziglina, I.N., Formation of planetesimals in the trans-Neptunian region of the protoplanetary disk, *Sol. Syst. Res.*, 2004, vol. 38, no. 4, pp. 288–300.

Marov, M.Ya., Dorofeeva, V.A., Ruskol, A.V., Kolesnichenko, A.V., Korolev, A.E., Samylkin, A.A., Makalkin, A.B., and Ziglina, I.N., Modeling of the for- mation and early evolution of preplanetary bodies, in *Problemy zarozhdeniya i evolyutsii biosfery (Origin and Evolution of Biosphere)*, Galimov, E.M., Ed., Moscow: Krasand, 2013, pp. 13–32.

Morbidelli, A., Bottke, W.F., Nesvorny, D., and Levison, H.F., Asteroids were born big, *Icarus*, 2009, vol. 204, pp. 558–573.

Myasnikov, V.P. and Titarenko, V.I., Evolution of self-gravitating clumps of a gas/dust nebula participating in the accumulation of planetary bodies, *Sol. Syst. Res.*, 1989, vol. 23, no. 1, p. 7.

Myasnikov, V.P. and Titarenko, V.I., Evolution of a self-gravitating gas/dust clump with allowance for radiative transfer in a diffusional approximation, *Sol. Syst. Res.*, 1990, vol. 23, no. 3, p. 126.

Nesvorny, D., Youdin, A.N., and Richardson, D.C., Formation of Kuiper belt binaries by gravitational collapse, *Astron. J.*, 2010, vol. 140, pp. 785–793.

Noll, K.S., Grundy, W.M., Stephens, D.C., Levison, H.F., and Kern, S.D., Evidence for two populations of classical transneptunian objects: the strong inclination dependence of classical binaries, *Icarus*, 2008a, vol. 194, pp. 758–768.

Noll, K.S., Grundy, W.M., Chiang, E.I., Margot, J.-L., and Kern, S.D., Binaries in the Kuiper belt, in *The Solar System beyond Neptune*, Barucci, M.A., Boehn- hardt, H., Cruikshank, D.P., and Morbidelli, A., Eds., Tucson: Univ. of Arizona Press, 2008b, pp. 345–363.

Petit, J.-M. and Mousis, O., KBO binaries: How numerous were they? *Icarus*, 2004, vol. 168, pp. 409–419.

Petit, J.-M., Kavelaars J.J., Gladman B.J., Margot J.L., Nicholson P.D., Jones R.L., Parker J.W., Ashby M.L.N., Campo Bagatin, A., Benavidez, P., Coffey, J., Rousselot, P., Mousis, O., and Taylor, P.A., The extreme Kuiper belt binary 2001 QW322, *Science*, 2008, vol. 322, pp. 432–434.

Poulet, F., Lucchetti, A., Bibring, J.P., Carter, J., Gondet, B., Jorda, L., Langevin, Y., Pilorget, C., Capanna, C., and Cremonese, G., Origin of the local structures at the Philae landing site and possible implications on the formation and evolution of 67P/Churyumov–Gerasi- menko, *Mon. Not. R. Astron. Soc.*, 2016, vol. 462, pp. S23–S32.

Pravec, P., Harris, A.W., and Warner, B.D., NEA rotations and binaries, Near-Earth objects, our celestial neighbors: opportunity and risk, *Proc. Int. Astronomical Union Symp.*, Milani, A., Valsecchi, G.B., and Vokrouhlicky, D., Eds., Cambridge: Cambridge Univ. Press, 2007, pp. 167–176.

Pravec, P., Scheirich, P., Vokrouhlický, D., Harris, A.W., Kusnirák, P., Hornoch, K., Pray, D.P., Higgins, D., Galád, A., Világi, J., et al., Binary asteroidal popula- tion. 2. Anisotropic distribution of orbit poles of small, inner main-belt binaries, *Icarus*, 2012, vol. 218, pp. 125–143.

Rein, H., Lesur, G., and Leinhardt, Z.M., The validity of the super-particle approximation during planetesimal formation, *Astron. Astrophys.*, 2010, vol. 511, art. ID A69, 9 p.

Richardson, D.R. and Walsh, K.J., Binary minor planets, *Annu. Rev. Earth Planet. Sci.*, 2006, vol. 34, pp. 47–81.

Rozitis, B., MacLennan, E., and Emery, J.P., Cohesive forces prevent the rotational breakup of rubble-pile asteroid (29075) 1950 DA, *Nature*, 2014, vol. 512, no. 7513, pp. 174–176.

Safronov, V.S., *Evolution of the Protoplanetary Cloud and Formation of the Earth and the Planets*, Washington, DC: Natl. Aeronaut. Space Admin., 1972, 212 p.

Sheppard, S.S., Ragozzine, D., and Trujillo, C., 2007 TY430: a cold classical Kuiper belt type binary in the Plutino population, *A s t r o n . J.*, 2012, vol. 143, no. 3, art ID. 58, 13 p

Stern, S.A., Collisional time scales in the Kuiper disk and their implications, *Astron. J.*, 1995, vol. 110, pp. 856–868.

Vityazev, A.V., Pechernikova, G.V., and Safronov, V.S., *Planety zemnoi gruppy: Proiskhozhdenie i rannyaya evolyutsiya (Terrestrial Planets: Origin and Early Evolution)*, Moscow: Nauka, 1990, 296 p.

Wahlberg Jansson, K. and Johansen, A., Formation of pebble-pile planetesimals, *Astron. Astrophys.*, 2014, vol. 570, art. ID A47, 11 p. https://arxiv.org/abs/1408.2535.

Wahlberg Jansson, K., Johansen, A., Bukhari, S.M., and Blum, J., The role of pebble fragmentation in planetesimal formation II. Numerical simulations, *Astrophys. J.*, 2017, vol. 835, art. ID 109, 11 p. https:// arxiv.org/abs/1609.07052.

Walsh, K.J., Richardson, D.R., and Michel, P., Rotational breakup as the origin of small binary asteroids, *Nature*, 2008, vol. 454, pp. 188–190.

Weidenschilling, S.J., On the origin of binary transneptunian objects, *Icarus*, 2002, vol. 160, pp. 212–215.

Weidenschilling, S.J., Radial drift of particles in the solar nebula: implications for planetesimal formation, *Icarus*, 2003, vol. 165, pp. 438–442.

Youdin, A.N., On the formation of planetesimals via secular gravitational instabilities with turbulent stirring, *Astrophys. J.*, 2011, vol. 731, art. ID A99, 18 p.

Youdin, A.N. and Kenyon, S.J., From disks to planets, in *Planets, Stars and Stellar Systems, Vol. 3: Solar and Stellar Planetary Systems*, Oswalt, T.D., French, L.M., and Kalas, P., Eds., Dordrecht: Springer-Verlag, 2013, pp. 1–62.

Ziglina, I.N. and Makalkin, A.B., Gravitational instability in the dust layer of a protoplanetary disk: Interaction of solid particles with turbulent gas in the layer, *Sol. Syst. Res.*, 2016, vol. 50, no. 6, pp. 408–425.